\documentclass{article}
\pdfoutput=1 
\usepackage[english]{babel}
\usepackage[T1]{fontenc}
\usepackage[a4paper, left=1in, right=1in, top = 1in, bottom = 1in]{geometry}
\usepackage{amsmath}

\usepackage{hyperref}
\hypersetup{
    colorlinks=true,
    linkcolor=black,
    filecolor=black,      
    urlcolor=black,
    citecolor =black,
}
\usepackage{cleveref}
\crefrangeformat{equation}{(#3#1#4)--(#5#2#6)}
\crefrangeformat{figure}{#3#1#4--#5#2#6}
\crefname{appendix}{}{}
\usepackage[flushleft]{threeparttable}
\usepackage{booktabs}
\usepackage{subcaption}
\newcommand\fnote[1]{\captionsetup{font=normalsize}\caption*{#1}}
\usepackage{lscape}
\usepackage{tgtermes}
\usepackage[percent]{overpic}
\usepackage{graphicx}
\usepackage{lipsum}
\usepackage{microtype}
\usepackage[usenames, dvipsnames]{color}
\usepackage{setspace}
\usepackage{enumitem}
\usepackage{subcaption}
\usepackage{tabularx}
\usepackage{multirow}
\usepackage[numbers,super]{natbib}
\makeatletter \renewcommand\@biblabel[1]{#1.} \makeatother
\usepackage[title]{appendix}
\usepackage{chngcntr}
\usepackage{authblk}

\title{Population-calibrated multiple imputation for a binary/categorical covariate in categorical regression models}
\author[1]{Tra My Pham\textsuperscript{$\star$}}
\author[2,3]{James R Carpenter}
\author[2]{Tim P Morris}
\author[4]{Angela M Wood}
\author[1]{Irene Petersen}
\affil[1]{Department of Primary Care and Population Health, University College London}
\affil[2]{London Hub for Trials Methodology Research, MRC Clinical Trials Unit at UCL}
\affil[3]{Department of Medical Statistics, London School of Hygiene and Tropical Medicine}
\affil[4]{Department of Public Health and Primary Care, University of Cambridge} 
\affil[$\star$] {tra.pham.09@ucl.ac.uk}

\date{}

\begin{document}

\maketitle
\begin{abstract}
Multiple imputation (MI) has become popular for analyses with missing data in medical research. The standard implementation of MI is based on the assumption of data being missing at random (MAR). However, for missing data generated by missing not at random (MNAR) mechanisms, MI performed assuming MAR might not be satisfactory. For an incomplete variable in a given dataset, its corresponding population marginal distribution might also be available in an external data source. We show how this information can be readily utilised in the imputation model to calibrate inference to the population, by incorporating an appropriately calculated offset termed the `calibrated-$\delta$ adjustment'. We describe the derivation of this offset from the population distribution of the incomplete variable and show how in applications it can be used to closely (and often exactly) match the post-imputation distribution to the population level. Through analytic and simulation studies, we show that our proposed calibrated-$\delta$ adjustment MI method can give the same inference as standard MI when data are MAR, and can produce more accurate inference under two general MNAR missingness mechanisms. The method is used to impute missing ethnicity data in a type 2 diabetes prevalence case study using UK primary care electronic health records, where it results in scientifically relevant changes in inference for non-White ethnic groups compared to standard MI. Calibrated-$\delta$ adjustment MI represents a pragmatic approach for utilising available population-level information in a sensitivity analysis to explore potential departure from the MAR assumption.    
\end{abstract}
\section{Introduction}\label{sec1}
Multiple imputation (MI) \cite{Rubin1987} has increasingly become a popular tool for analyses with missing data in  medical research \cite{Sterne2009, Klebanoff2008}; the method is now incorporated in many standard statistical software packages.\cite{StataCorp2015a, VanBuuren2011, Yuan2011} In MI, several completed datasets are created and in each, missing data are replaced with values drawn from an imputation model which is the Bayesian posterior predictive distribution of the missing data, given the observed data. Each completed dataset is then analysed using the substantive analysis model that would have been used had there been no missing data. This process generates several sets of parameter estimates, which are then combined into a single set of results using Rubin's rules.\cite{Rubin1987, Barnard1999} Given congenial specification of the imputation model, Rubin's rules provide estimates of standard errors and confidence intervals that correctly reflect the uncertainty introduced by missing data.
\par
The standard implementation of MI in widely available software packages provides valid inference under the assumption that missing values are missing completely at random (MCAR) or missing at random (MAR). However, in many applied settings, it is possible that the unseen data are missing not at random (MNAR). For example, in primary care, individuals with more frequent blood pressure readings may, on average, have higher blood pressure compared to the rest of the primary care population. Although MI can be used when data are MNAR, imputation becomes more difficult because a model for the missing data mechanism needs to be specified, which describes how missingness depends on both observed and unobserved quantities. This implies that in practice, it is necessary to define a model for either the association between the probability of observing a variable and its unseen values (selection models) \cite{Little-Rubin2002}; or the difference in the distribution of subjects with and without missing data (pattern-mixture models).\cite{Little1993, Little1994} Due to the potential complexity of modelling the missingness mechanism under MNAR, analyses assuming MNAR are relatively infrequently performed and reported in the applied literature. Instead, in practice, researchers more often try to enhance the plausibility of the MAR assumption as much as possible by including many variables in the imputation model.\cite{White2011, Collins2001}
\par
%
The extra model specification requirement in MI for MNAR data raises several issues. Firstly, the underlying MAR and MNAR mechanisms are not verifiable from the observed data alone. Secondly, there can be an infinite number of possible MNAR models for any dataset, and it is very rare to know which of these models is appropriate for the missingness mechanism. However, for an incomplete variable in a given dataset, its corresponding population marginal distribution might be available from an external data source, such as a population census or survey. If our study sample in truth comes from such a population, it is sensible to feed this population information into the imputation model, in order to calibrate inference to the population. 
\par
In this paper, we propose a version of MI for an incomplete binary/categorical variable, termed \textit{calibrated-$\delta$ adjustment MI}, which exploits such external information. In this approach, the population distribution of the incomplete variable can be used to calculate an adjustment in the imputation model's intercept, which is used in MI such that the post-imputation distribution much more closely (and often exactly) matches the population distribution. The idea of the calibrated-$\delta$ adjustment is motivated by van Buuren et al.'s $\delta$ adjustment (offset) approach in MI.\cite{VanBuuren1999} However, while values of $\delta$ are often chosen arbitrarily (and independently of covariates in the imputation model) in van Buuren et al.'s approach, the incomplete variable's population distribution is used to derive the value of $\delta$ in calibrated-$\delta$ adjustment MI. We show that our proposed method gives equivalent inference to standard MI when data are MAR, and can produce unbiased inference under two general MNAR mechanisms.
\par
From a practical point of view, the development of calibrated-$\delta$ adjustment MI is motivated by the issue incomplete recording of ethnicity data in UK primary care electronic health records. Routine recording of ethnicity has been incorporated at the general practice level in the UK, and the variable is therefore available in many large primary care databases. However, research addressing ethnicity has been constrained by the low level of recording.\cite{Kumarapeli2006, Aspinall2007, Mathur2013b} Studies often handle missing data in ethnicity by either dropping ethnicity from the analysis \cite{Osborn2015}, performing a complete record analysis (i.e. excluding individuals with missing data), or single imputation of missing values with the White ethnic group \cite{Hippisley-Cox2008}; these methods will generally lead to biased estimates of association and standard errors.\cite{Sterne2009} In addition, the probability that ethnicity is recorded in primary care may well vary systematically by ethnic groups, even after adjusting for other variables.\cite{Mathur2013b} This implies a potential MNAR mechanism for ethnicity, and as a result, standard MI might fail to give valid inference for the underlying population. Since the population marginal distribution of ethnicity is available in the UK census data, the plausibility of the MAR assumption for ethnicity in UK primary care data can be assessed by using standard MI to handle missing data, and comparing the resulting ethnicity distribution to that in the census. In earlier work, we explored departures from the MAR assumption for other incomplete heath indicators by comparing the results with external nationally representative datasets.\cite{Marston2010, Marston2014} As an example of this, Marston et al. (2014) reported that if smoking status is missing for a patient then he or she is typically either an ex-smoker or non-smoker, and accordingly proposed only allowing imputed data to take one of these two values \cite{Marston2014}. The method we describe here supersedes this ad-hoc approach, providing a way to incorporate population distribution information into MI.
\par
This paper focuses on missing data in an incomplete binary/categorical covariate in an analysis model, where the outcome variable and other covariates are all binary/categorical and fully observed. The remainder of this paper is structured as follows. \Cref{sec2} works through a simple example analytically to describe the derivation of the calibrated-$\delta$ adjustment. In \cref{sec3}, we formally introduce the procedure of calibrated-$\delta$ adjustment MI and evaluate the performance of the method in simulation studies. \Cref{sec4} illustrates the use of this MI method in a case study which uses electronic health records to examine the association between ethnicity and the prevalence of type 2 diabetes diagnoses in UK primary care. We conclude the paper with a discussion in \cref{sec5}.

\section{Analytic study -- bias in a $2\times 2$ contingency table}
\label{sec2}
In this section, we present the development of calibrated-$\delta$ adjustment MI in a simple setting of a $2 \times 2$ contingency table and describe the derivation of the calibrated-$\delta$ adjustment.
\par
Suppose it is of interest to study the association between a binary variable $x$ taking values $j=0,1$ and a binary outcome $y$ taking values $k=0,1$, whose full-data distribution is given in Table \ref{tab:analytic_study_full_data}. The full-data distribution is assumed to be identical to the population distribution, such that the population marginal distribution of $x$ is given by $p_{j}^{\text{pop}} = \frac{n_{j+}}{n_{++}}$. The data generating model is
\begin{equation*}  
        \text{logit}\left[p\left(y=1\mid x\right)\right] = \beta_{0} + \beta_{x}x,
\end{equation*}
whose parameters can be written in terms of cell counts, $\beta_{0} = \text{ln}\left(\frac{n_{01}}{n_{00}}\right)$ and $\beta_{x} = \text{ln}\left(\frac{n_{11}n_{00}}{n_{01}n_{10}}\right)$. 
\begin{table}[b!]
	\renewcommand{\arraystretch}{1.1}
	    \centering
        \caption{Analytic study: distribution of $x$ and $y$ and selection models for missingness in $x$.}
        \begin{subtable}[t]{\linewidth}
        \centering
        \subcaption{Distribution in the full data of size $n$.}
        \label{tab:analytic_study_full_data}
        \begin{tabular}{lccc}         
        \toprule
                   & $y=0$       & $y=1$    & $\sum_{j=0}^{1}x$ \\ \midrule
        $x=0$      & $n_{00}$    & $n_{01}$ & $n_{0+}$   \\
        $x=1$      & $n_{10}$    & $n_{11}$ & $n_{1+}$   \\ \midrule
        $\sum_{k=0}^{1} y$ & $n_{+0}$ & $n_{+1}$ & $n_{++}$           \\ \bottomrule
        \end{tabular} 
        \end{subtable}
        \vskip 15pt
        \begin{subtable}[t]{\linewidth}
        \centering
        \subcaption{Distribution among subjects with observed $x$ ($y$ is fully observed).}
        \label{tab:analytic_study_r=1}
        \centering
        \begin{tabular}{lcccc}
        \toprule
                             & $y=0 \mid r=1$                 & $y=1 \mid r=1$        & $\sum_{j=0}^{1} x\mid r=1$  & Population       \\ \midrule
        $x=0 \mid r=1$                & $n_{00}^{\text{obs}}$ & $n_{01}^{\text{obs}}$ & $n_{0+}^{\text{obs}}$ & $n_{0+}$         \\
        $x=1 \mid r=1$                & $n_{10}^{\text{obs}}$ & $n_{11}^{\text{obs}}$ & $n_{1+}^{\text{obs}}$ & $n_{1+}$         \\ \midrule
        $\sum_{k=0}^{1}y\mid r=1$ & $n_{+0}^{\text{obs}}$ & $n_{+1}^{\text{obs}}$ & $n_{++}^{\text{obs}}$ &                  \\
        $\sum_{k=0}^{1}y\mid r=0$ & $n_{+0}^{\text{mis}}$ & $n_{+1}^{\text{mis}}$ & $n_{++}^{\text{mis}}$ &                  \\ \bottomrule
        \end{tabular}
        \end{subtable}
        \vskip 15pt
        \begin{subtable}[t]{\linewidth}
    	\centering
    	\subcaption{Models for missingness in $x$.}
    	\label{tab:analytic_study_selections}
	    \begin{tabular}{lcc}
	    \toprule
	    \begin{tabular}[l]{@{}l@{}}Linear predictor of selection model\\ $\text{logit}\left[p\left[(r=1 \mid x, y\right)\right]$\end{tabular} & \begin{tabular}[c]{@{}c@{}}Selection probability\\ $p\left(r_{jk} = 1\right)$\end{tabular} & \begin{tabular}[c]{@{}c@{}}Label\end{tabular} \\ \midrule
$\alpha_{0}$       & $p_{r}$    & M1    \\
$\alpha_{0} + \alpha_{y}y$    & $p_{r_{k}}$   & M2    \\
$\alpha_{0} + \alpha_{x}x$    & $p_{r_{j}}$    & M3    \\
$\alpha_{0} + \alpha_{x}x + \alpha_{y}y$   & $p_{r_{jk}}$   & M4    \\ \bottomrule
	    \end{tabular}
        \end{subtable}
        \\
        \fnote{Note: $r$: response indicator of $x$; $j$ and $k$: index categories of $x$ and $y$, respectively; $j, k$ take values $0/1$.}
\end{table}
\par
In addition, suppose that $y$ is fully observed, while some data in $x$ are set to missing (i.e. the sample contains no individuals with missing $y$ and observed $x$, Table \ref{tab:analytic_study_r=1}). Let $r$ be the response indicator taking values $1$ if $x$ is observed and $0$ if $x$ is missing. Four different missingness mechanisms considered for $x$ and the corresponding selection models are presented in Table \ref{tab:analytic_study_selections}. Observed cell counts, $n_{jk}^{\text{obs}}$, can be written as a product of the full-data cell counts, $n_{jk}$, and the cell-wise probability of observing $x$, $p_{r_{jk}}$, such that $n_{jk}^{\text{obs}} = n_{jk}p_{r_{jk}}$. 
\par
To perform standard MI of missing values in $x$, an imputation model
\begin{equation}
    \label{eq:standard_imp_model} 
        \text{logit}\left[p\left(x=1\mid y\right)\right] = \theta_{0} + \theta_{y}y,
\end{equation}
is fitted to the $n_{++}^{\text{obs}}$ complete records (Table \ref{tab:analytic_study_r=1}) to obtain the $\theta$ parameter estimates, where
\begin{equation*}
    \theta_{0}^{\text{obs}} = \text{ln}\left(\frac{n_{10}^{\text{obs}}}{n_{00}^{\text{obs}}}\right); \quad \theta_{y}^{\text{obs}} = \text{ln}\left(\frac{n_{11}^{\text{obs}}n_{00}^{\text{obs}}}{n_{01}^{\text{obs}}n_{10}^{\text{obs}}}\right).
\end{equation*}
When $x$ is MCAR or MAR conditional on $y$, we can obtain an unbiased estimate of the association between $x$ and $y$ in the missing data by fitting the above logistic regression imputation model to the complete records. No adjustment is needed in the intercept of the imputation model, and standard MI provides unbiased estimates of the marginal distribution of $x$ as well as the association between $x$ and $y$. We focus on two general MNAR mechanisms described below.
%
\subsection{$x$ is MNAR dependent on $x$}
%
    Under this missingness mechanism, the posited model for the response indicator $r$ of $x$ is given by
    \begin{equation}
        \label{eq:selection_mnar_x}
        \text{logit}\left[p\left(r=1\mid x\right)\right] = \alpha_{0} + \alpha_{x}x,
    \end{equation}
    and the corresponding probabilities of observing $x$ are
    \begin{equation*}
        p\left(r=1 \mid x=j\right) = p_{r_{j}} = \text{expit}\left(\alpha_{0} + \alpha_{x}x\right); \quad j=0,1.
    \end{equation*}    
    For imputation model \eqref{eq:standard_imp_model}, the log odds ratios of $x=1$ for $y=1$ compared to $y=0$ in the observed and missing data are 
    \begin{align*}
        \left[\theta_{y} \mid r=1\right] &= \theta_{y}^{\text{obs}} = \text{ln}\left(\frac{n_{00}p_{r_{0}}n_{11}p_{r_{1}}}{n_{01}p_{r_{0}}n_{10}p_{r_{1}}}\right) = \text{ln}\left(\frac{n_{00}n_{11}}{n_{01}n_{10}}\right); \\
        \left[\theta_{y} \mid r=0\right] &= \theta_{y}^{\text{mis}} = \text{ln}\left(\frac{n_{00}\left(1-p_{r_{0}}\right)n_{11}\left(1-p_{r_{1}}\right)}{n_{01}\left(1-p_{r_{0}}\right)n_{10}\left(1-p_{r_{1}}\right)}\right) = \text{ln}\left(\frac{n_{00}n_{11}}{n_{01}n_{10}}\right),
    \end{align*}
    respectively. Hence, $\theta_{y}^{\text{obs}} = \theta_{y}^{\text{mis}}$, which are also the same as the log odds ratio $\theta_{y}$ in the full data (i.e. before values in $x$ are set to missing). The log odds of $x=1$ for $y=0$ in the observed and missing data are given by   
    \begin{align*}
        \left[\theta_{0} \mid r=1\right] &= \theta_{0}^{\text{obs}} = \text{ln}\left(\frac{n_{10}p_{r_{1}}}{n_{00}p_{r_{0}}}\right); \\
        \left[\theta_{0} \mid r=0\right] &= \theta_{0}^{\text{mis}} = \text{ln}\left(\frac{n_{10}\left(1-p_{r_{1}}\right)}{n_{00}\left(1-p_{r_{0}}\right)}\right),
    \end{align*}
    respectively. This implies that the correct adjustment in the imputation model's intercept
    should be
    \begin{align*}
        \theta_{0}^{\text{mis}} - \theta_{0}^{\text{obs}} &= \text{ln}\left(\frac{\left(1-p_{r_{1}}\right)p_{r_{0}}}{\left(1-p_{r_{0}}\right)p_{r_{1}}}\right) \\
        &= \text{ln}\left(\frac{\text{exp}\left(\alpha_{0}\right)}{\text{exp}\left(\alpha_{0}+\alpha_{x}\right)}\right) \\
        &= -\alpha_{x},
    \end{align*}
    which is minus the log odds ratio of observing $x$ for $x = 1$ compared to $x = 0$ in \eqref{eq:selection_mnar_x}.
%
\subsection{$x$ is MNAR dependent on $x$ and $y$}
%
    Under this missingness mechanism, the posited model for the response indicator $r$ of $x$ is given by
    \begin{equation}
        \label{eq:selection_mnar_x_y}
        \text{logit}\left[p\left(r=1 \mid x,y\right)\right] = \alpha_{0} + \alpha_{x}x + \alpha_{y}y,
    \end{equation}
    and the corresponding probabilities of observing $x$ are
    \begin{equation*}
        p\left(r=1 \mid x=j, y=k\right) = p_{r_{jk}} = \text{expit}\left(\alpha_{0} + \alpha_{x}x + \alpha_{y}y\right); \quad j,k = 0,1.
    \end{equation*}
    For imputation model \eqref{eq:standard_imp_model}, the log odds ratios of $x=1$ for $y=1$ compared to $y=0$ in the observed and missing data are
    \begin{align}
        \theta_{y}^{\text{obs}} &= \text{ln}\left(\frac{n_{00}p_{r_{00}}n_{11}p_{r_{11}}}{n_{01}p_{r_{01}}n_{10}p_{r_{10}}}\right); \label{eq:thetay_obs_mnar_x_y}\\
        \theta_{y}^{\text{mis}} &= \text{ln}\left(\frac{n_{00}\left(1-p_{r_{00}}\right)n_{11}\left(1-p_{r_{11}}\right)}{n_{01}\left(1-p_{r_{01}}\right)n_{10}\left(1-p_{r_{10}}\right)}\right). \label{eq:thetay_mis_mnar_x_y}       
    \end{align}
    Again, it can be shown from \eqref{eq:thetay_obs_mnar_x_y} and \eqref{eq:thetay_mis_mnar_x_y} that $\theta_{y}^{\text{obs}} = \theta_{y}^{\text{mis}}$, since   
    \begin{align*}
    \theta_{y}^{\text{mis}} - \theta_{y}^{\text{obs}} &= \text{ln}\left(\frac{\left(1-p_{r_{00}}\right)\left(1-p_{r_{11}}\right)p_{r_{01}}p_{r_{10}}}{\left(1-p_{r_{01}}\right)\left(1-p_{r_{10}}\right)p_{r_{00}}p_{r_{11}}}\right)\\
    &= \text{ln}\left(\frac{\text{exp}\left(\alpha_{0} + \alpha_{x}\right)\text{exp}\left(\alpha_{0} + \alpha_{y}\right)}{\text{exp}\left(\alpha_{0}\right)\text{exp}\left(\alpha_{0} + \alpha_{x} + \alpha_{y}\right)}\right) \\
    &= 0.
    \end{align*}
    The log odds of $x=1$ for $y=0$ in the observed and missing data are given by
    \begin{align*}
        \theta_{0}^{\text{obs}} = \text{ln}\left(\frac{n_{10}p_{r_{10}}}{n_{00}p_{r_{00}}}\right); \\
        \theta_{0}^{\text{mis}} = \text{ln}\left(\frac{n_{10}\left(1-p_{r_{10}}\right)}{n_{00}\left(1-p_{r_{00}}\right)}\right),
    \end{align*}
    which implies that the correct adjustment in the imputation model's intercept should be
    \begin{align*}
        \theta_{0}^{\text{mis}} - \theta_{0}^{\text{obs}} &= \text{ln}\left(\frac{\left(1-p_{r_{10}}\right)p_{r_{00}}}{\left(1-p_{r_{00}}\right)p_{r_{10}}}\right) \\
        &= \text{ln}\left(\frac{\text{exp}\left(\alpha_{0}\right)}{\text{exp}\left(\alpha_{0}+\alpha_{x}\right)}\right) \\
        &= -\alpha_{x},
    \end{align*}
    which is again minus the log odds ratio of observing $x$ in \eqref{eq:selection_mnar_x_y}.
%
\subsection{Derivation of the calibrated-$\delta$ adjustment}
%
The analytic calculations above confirm that in a $2\times 2$ contingency table setting, appropriately adjusting the intercept of the imputation model for the covariate $x$ can sufficiently correct bias introduced by MNAR mechanisms under which missingness in $x$ depends on either its values or both its values and the outcome (M3 and M4). The population distribution of $x$ can be used to calculate the correct adjustment in the imputation model's intercept. This adjustment is referred to as the \textit{calibrated-$\delta$ adjustment} to clarify its relationship to van Buuren et al.'s $\delta$ adjustment.\cite{VanBuuren1999}
\par
The probability of $x=1$ can be written in terms of the conditional probabilities among subjects with observed and missing $x$
\begin{equation*}
    p\left(x=1\right) = p\left(x=1 \mid r=1\right)p\left(r=1\right) + p\left(x=1 \mid r=0\right)p\left(r=0\right),
\end{equation*}
where $p\left(x=1\right)$ is the population proportion; $p\left(x=1 \mid r=1\right)$ , $p\left(r=1\right)$, and $p\left(r=0\right)$ can be obtained from the observed data. Thus, $p\left(x=1 \mid r=0\right)$ can be solved for as
\begin{equation}
    \label{eq:p_x=1_partitioned2}
    p\left(x=1 \mid r=0\right) = \frac{p\left(x=1\right) - p\left(x=1 \mid r=1\right)p\left(r=1\right)}{p\left(r=0\right)}.
\end{equation}
Note that $p\left(x=1 \mid r=0\right)$ can be further written as
\begin{align}
    p\left(x=1 \mid r=0\right) &= \sum_{k=0}^{1}p\left(x=1 \mid y=k, r=0\right)p\left(y=k \mid r=0\right) \nonumber\\
    &= \sum_{k=0}^{1}\text{expit}\left(\theta_{0}^{\text{mis}} + \theta_{y}^{\text{mis}} I\left[y=k\right]\right)\frac{n_{+k}^{\text{mis}}}{n_{++}^{\text{mis}}} \nonumber\\
    &= \frac{1}{n_{++}^{\text{mis}}}\text{expit}\left(\theta_{0}^{\text{mis}} + \theta_{y}^{\text{mis}} I\left[y=k\right]\right)n_{+k}^{\text{mis}}, \label{eq:p_x=1_r=0}
\end{align}
where $I\left[A\right]$ is an indicator function taking values 1 if $A$ is true and 0 otherwise. It is shown earlier that when $x$ is MNAR dependent on either the values of $x$ or both $x$ and $y$, $\theta_{y}^{\text{obs}}=\theta_{y}^{\text{mis}}$; \eqref{eq:p_x=1_r=0} is therefore equal to
\begin{align*}
    p\left(x=1 \mid r=0\right) &= \frac{1}{n_{++}^{\text{mis}}}\text{expit}\left(\theta_{0}^{\text{mis}} + \theta_{y}^{\text{obs}} I\left[y=k\right]\right)n_{+k}^{\text{mis}} \\
    &= \frac{1}{n_{++}^{\text{mis}}}\text{expit}\left(\left(\theta_{0}^{\text{obs}} + \delta\right) + \theta_{y}^{\text{obs}} I\left[y=k\right]\right)n_{+k}^{\text{mis}} \\
    &= \frac{1}{n^{\text{mis}}}\sum_{i=1}^{n^{\text{mis}}}\text{expit}\left(\left(\theta_{0}^{\text{obs}} + \delta \right) + \theta_{y}^{\text{obs}}y_{i}\right),
\end{align*}
where $\delta$ is the adjustment factor in the intercept of the imputation model for $x$. The value of the calibrated-$\delta$ adjustment can be obtained numerically from \eqref{eq:p_x=1_partitioned2} and \eqref{eq:p_x=1_r=0} using interval bisection \cite{Russ1980,Burden2011} (or any other root-finding method).
\par
When the population marginal distribution of the incomplete covariate $x$ is available, a natural alternative to adjusting the intercept of the imputation model based on this information is to weight the complete records in the imputation model (which we term `weighted multiple imputation'), in order to match the post-imputation distribution of $x$ to the population. In the supporting information section we explore two such weighting approaches, marginal and conditional weighted MI; we show analytically that while these methods can provide more accurate results compared to standard MI under certain MNAR mechanisms, they do not provide a general solution as does calibrated-$\delta$ adjustment MI.  

\section{Simulation studies}\label{sec3}
This section presents univariate simulation studies to evaluate performance measures of the calibrated-$\delta$ adjustment MI method for an incomplete binary covariate $x$, when the fully observed outcome variable $y$ is also binary. The term `univariate' is used here to refer to the setting where missingness occurs in a single covariate. The aims of these simulation studies are (i) to examine finite-sample properties of calibrated-$\delta$ adjustment MI including bias in parameter estimates, efficiency in terms of the empirical and average model standard errors (SE), and coverage of $95\%$ confidence intervals (CI); and (ii) to compare the method with standard MI and complete record analysis (CRA) under various missingness mechanisms for $x$.

\subsection{When the population distribution is `known'}
\label{subsec3.1}
Below we consider the setting where the population distribution of the incomplete variable is obtained from a population census or equivalent, i.e. it is `known'. The uncertainty associated with having to estimate the population distribution is explored in \cref{subsec3.2}.

\subsubsection{Method}
\label{subsubsec3.1.1}
Similar to the analytic study presented in \cref{sec2}, the analysis model in this simulation study is a logistic regression model for a fully observed binary outcome $y$ on an incomplete binary covariate $x$. Calibrated-$\delta$ adjustment MI is compared to standard MI and CRA under four missingness mechanisms of increase complexity. The data generating mechanism and analysis procedures are as follows.
\begin{enumerate}
    \item Simulate $n=5\,000$ complete values of the binary $0/1$ covariate $x$ and binary $0/1$ outcome $y$ from the following models
    \begin{align}
        &x \sim \text{Bernoulli}\left(p_{x}^{\text{pop}} = 0.7\right); \nonumber\\
        & \text{logit}\left[p\left(y=1 \mid x\right)\right] = \beta_{0} + \beta_{x}x, \label{eq:moi_sim1}
    \end{align}
    where $\beta_{0}$ and $\beta_{x}$ are arbitrarily set to $\text{ln}\left(0.5\right)$ and $\text{ln}\left(1.5\right)$, respectively. The same values of the $\beta$ parameters are used throughout to make bias comparable across all simulation settings. This sample size is chosen to minimise the issue of small-sample bias associated with the logistic regression \cite{Nemes2009};
    \item Simulate a binary indicator of response $r$ of $x$ from each of the selection models M1--M4 (Table \ref{tab:analytic_study_selections}). Values of $1.5$ and $-1.5$ are chosen for $\alpha_{y}$ and $\alpha_{x}$ in M2 and M3, respectively, to reflect strong odds ratios (OR) of observing $x$ (OR $= 4.5$ and $0.2$, respectively). For M4, $\alpha_{y} = 1.5$ and $\alpha_{x} = -1.5$ are chosen as bias in the three MI methods under evaluation is likely to be apparent with these coefficients predicting missingness in $x$. For all selection models, $\alpha_{0}$ is altered to achieve approximately $45\%$ missing $x$. For M1, $\alpha_{0}$ is calculated directly as $\text{ln}\left(\frac{0.55}{0.45}\right)$; for M2--M4, $\alpha_{0} = -0.2; 1.35$; and $0.75$ appear to work well;
    \item For $i=1 \ldots 5\,000$, set $x_{i}$ to missing if $r_{i}=0$;
    \item Impute missing values in $x$ $M=50$ times using standard MI and calibrated-$\delta$ adjustment MI in turn;
    \item In each MI method, fit the analysis model \eqref{eq:moi_sim1} to each completed dataset and combine the results using Rubin's rules.\cite{Rubin1987, Barnard1999}
\end{enumerate}
Steps 1--5 are repeated $S=2\,000$ times under each of the four selection models M1--M4, so the same set of simulated independent datasets is used to compare the three MI methods under the same missingness scenario, but a different set of datasets is generated for each missingness scenario.\cite{Burton2006} The parameters of interest are $\beta_{0}$ and $\beta_{x}$, although in practice $\beta_{x}$ is usually of more interest. Bias, efficiency of $\hat{\beta}_{0}$ and $\hat{\beta}_{x}$ in terms of the empirical standard errors, and coverage of 95\% CIs are calculated over $2\,000$ repetitions for each combination of simulation settings,\cite{White2010a} with analyses of full data (i.e. before any values in $x$ are set to missing) and complete records also provided for comparison. 
\par
All simulations are performed in Stata 14 \cite{StataCorp2015b}; \texttt{mi impute logit} is used for standard MI, the community-contributed command \texttt{uvis logit} \cite{Royston2004} for calibrated-$\delta$ adjustment MI, and \texttt{mi estimate: logit} for fitting the analysis model to the completed datasets and combining the results using Rubin's rules.\cite{Rubin1987, Barnard1999} Simulated datasets are analysed using the community-contributed command \texttt{simsum}.\cite{White2010a}
\par
Based on the analytic calculations presented in \cref{sec2}, we propose the following procedure for imputing missing values in the covariate $x$ using calibrated-$\delta$ adjustment MI.
\begin{enumerate}
    \item Fit a logistic regression imputation model for $x$ conditional on $y$ to the complete records to obtain the maximum likelihood estimates of the imputation models' parameters $\hat{\theta}$ and their asymptotic sampling variance $\widehat{\boldsymbol{U}}$;
    \item Draw new parameters $\tilde{\boldsymbol{\theta}}$ from the large-sample normal approximation $N(\widehat{\boldsymbol{\theta}}, \widehat{\boldsymbol{U}})$ of their posterior distribution, assuming non-informative priors;
    \item Draw a new probability of observing $x$, $\tilde{p}_{r}$, from the normal approximation $N\left(\hat{p}_{r}, \frac{\hat{p}_{r}\left(1-\hat{p}_{r}\right)}{n}\right)$, where $\hat{p}_{r}$ is the sample proportion of the response indicator of $x$, $\hat{p}_{r} = \frac{n_{++}^{\text{obs}}}{n_{++}}$;
    \item Draw a new probability of observed $x=1$, $\tilde{p}_{x}$, from the normal approximation $N\left(\hat{p}_{x}, \frac{\hat{p}_{x}\left(1-\hat{p}_{x}\right)}{n}\right)$, where $\hat{p}_{x}$ is the observed proportion of $x=1$, $\hat{p}_{x} = \frac{n_{1+}^{\text{obs}}}{n_{++}^{\text{obs}}}$;
    \item Derive the value of the calibrated-$\delta$ adjustment from the equation
    \begin{equation*}
        \frac{1}{n^{\text{mis}}}\sum_{i=1}^{n^{\text{mis}}}\text{expit}\left(\left(\tilde{\theta}_{0} + \delta\right) + \tilde{\theta}_{y}y_{i}\right) = \frac{p_{x}^{\text{pop}} - \tilde{p}_{x}}{\tilde{p}_{r}},
    \end{equation*}
    where $p_{x}^{\text{pop}}$ is the probability of $x=1$ in the population;
    \item Fit the logistic regression imputation model for $x$ conditional on $y$ (in step 1) to the complete records with the intercept adjustment fixed to $\delta$ to obtain the maximum likelihood estimates of the imputation models' parameters $\hat{\boldsymbol{\theta}}$ and their asymptotic sampling variance $\widehat{\boldsymbol{U}}$;
    \item Draw new parameters $\dot{\boldsymbol{\theta}}$ from the large-sample normal approximation $N(\widehat{\boldsymbol{\theta}}, \widehat{\boldsymbol{U}})$ of their posterior distribution, assuming non-informative priors;  
    \item Draw imputed values for $x$ from the above logistic regression imputation model, using the newly drawn parameters $\dot{\boldsymbol{\theta}}$ and calibrated-$\delta$ adjustment.
\end{enumerate}

\subsubsection{Results}
\label{subsubsec3.1.2}
\begin{figure}[htbp!]
    \centering
    \caption{Simulation study: bias in point estimates, empirical and average model SE, and coverage of 95\% CIs under different missingness mechanism for $x$.}
    \label{fig:sim_base}
    \begin{subfigure}[b]{0.69\textwidth}
        \includegraphics[width=\textwidth]{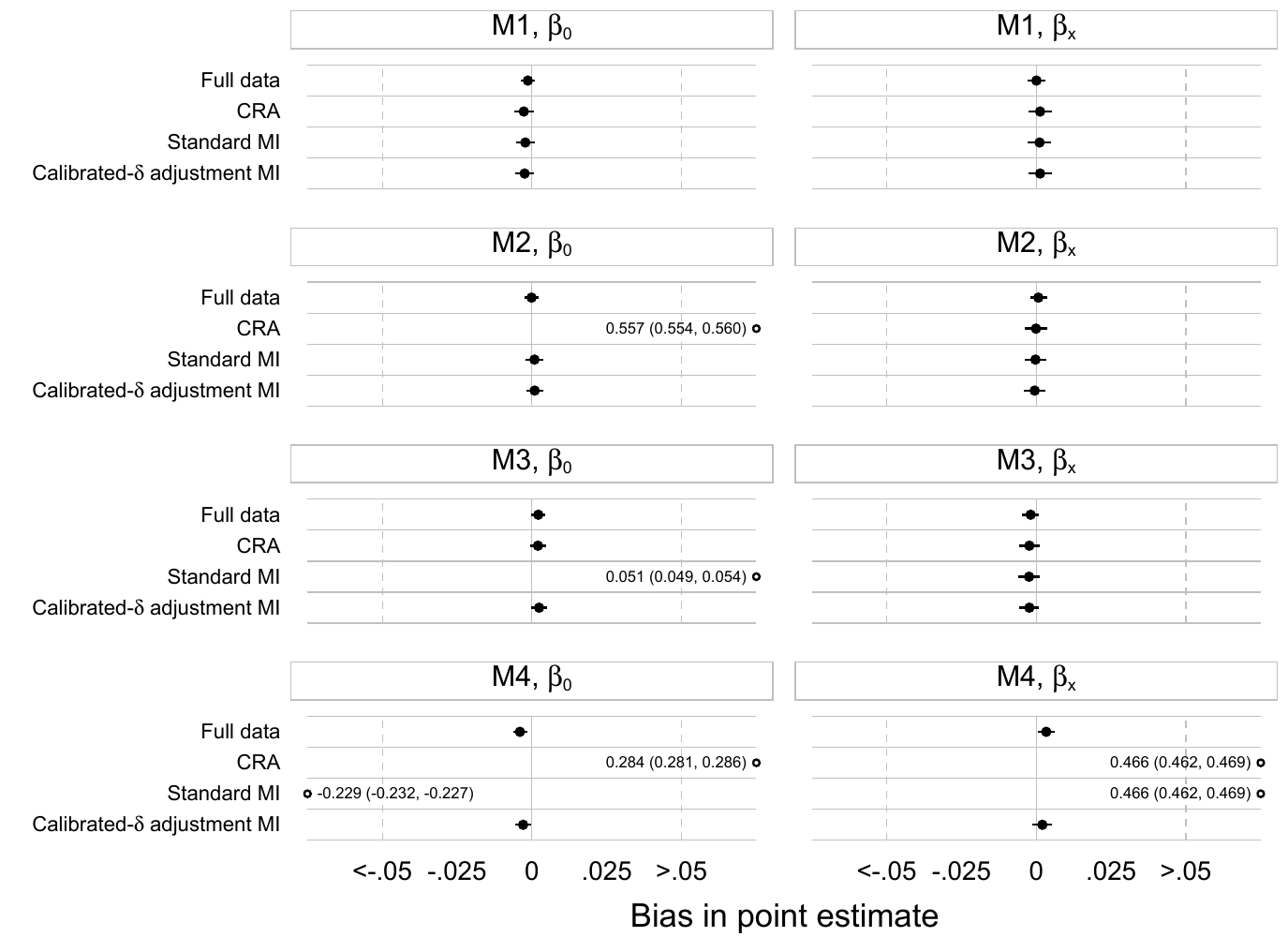}
    \end{subfigure}
    ~ 
    \begin{subfigure}[b]{0.69\textwidth}
        \includegraphics[width=\textwidth]{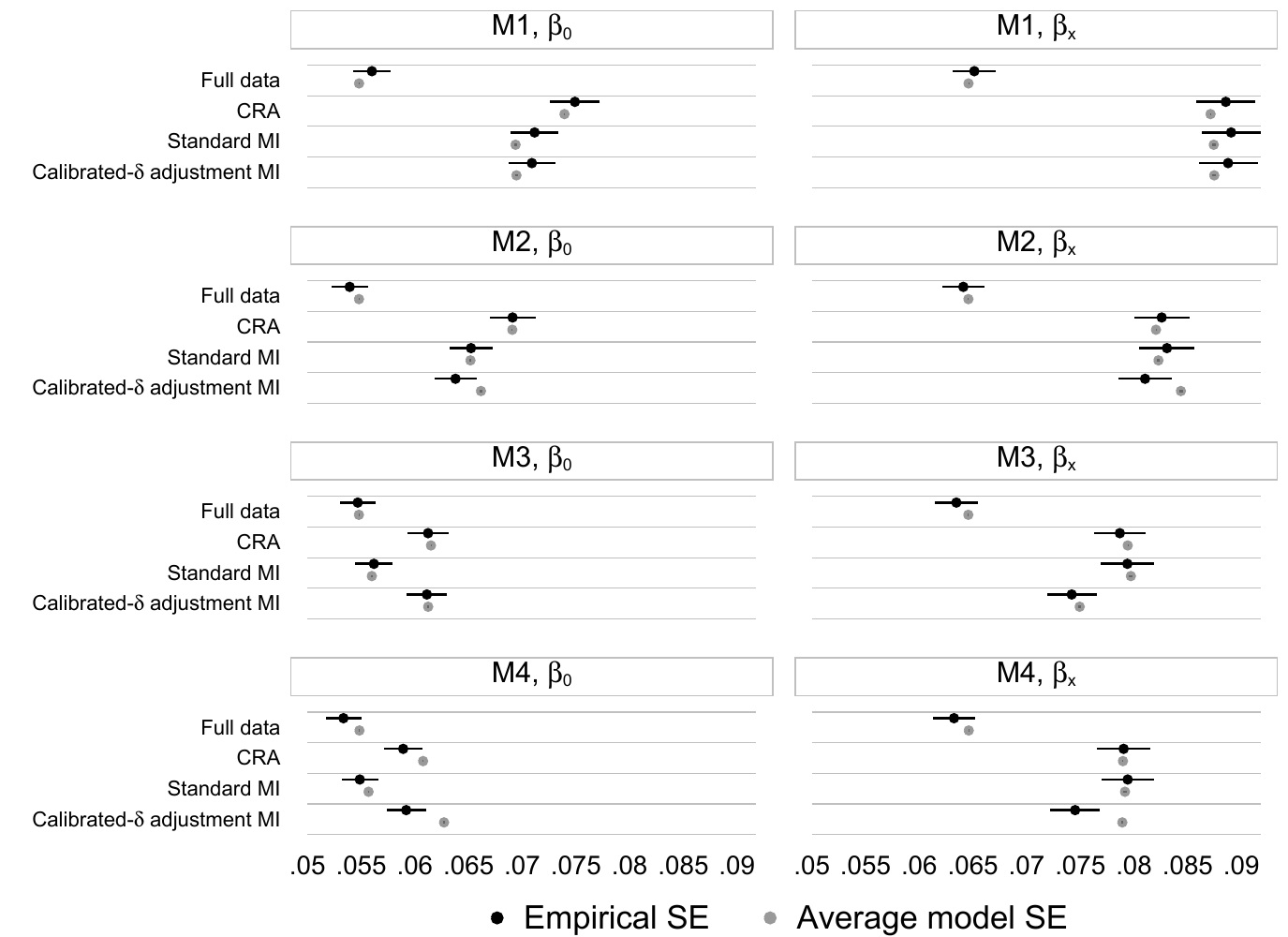}
    \end{subfigure}
    ~ 
    \begin{subfigure}[b]{0.69\textwidth}
        \includegraphics[width=\textwidth]{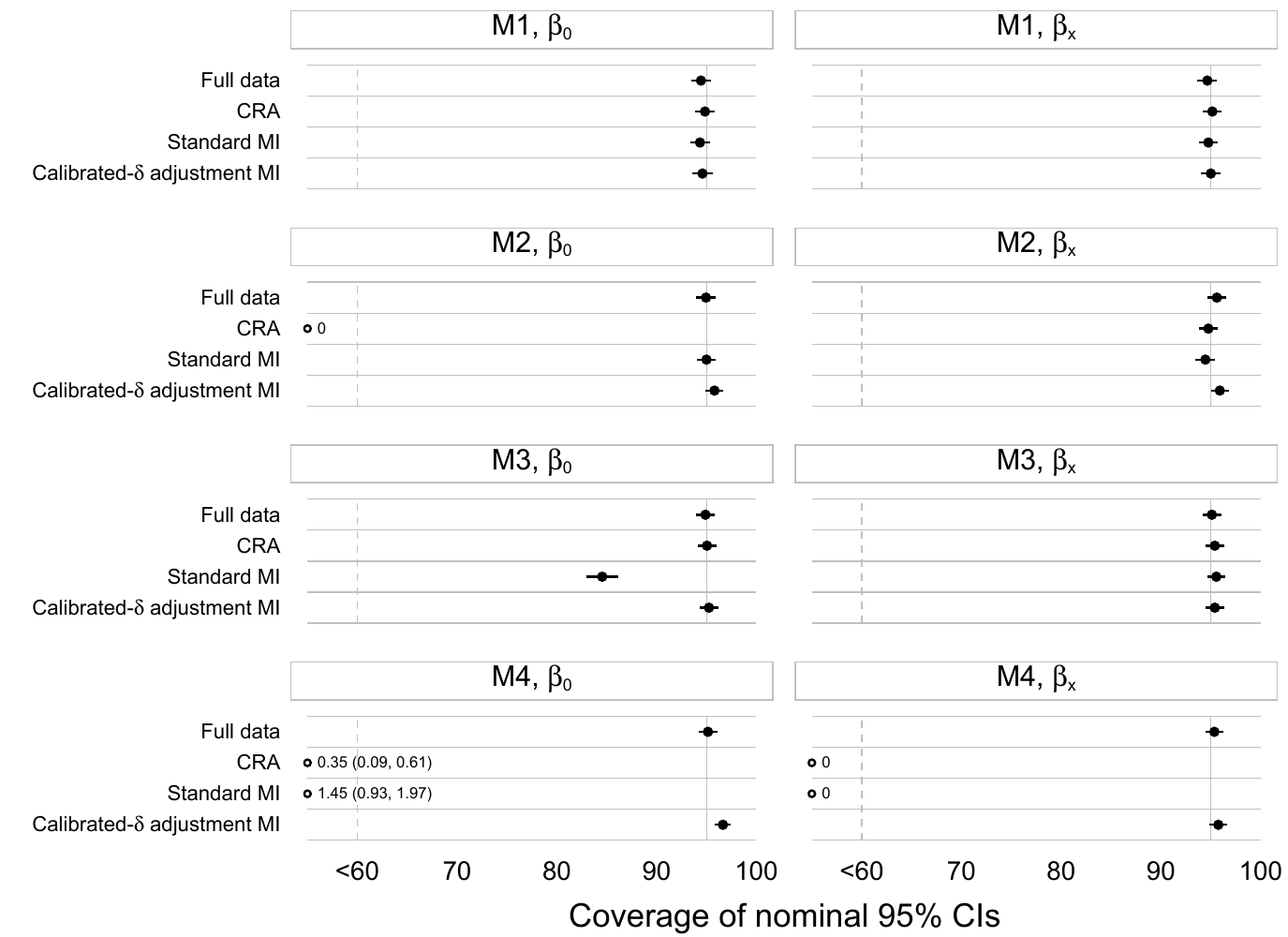}
    \end{subfigure}
\end{figure}

Results of the simulation study are summarised graphically in Figure \ref{fig:sim_base}. Full data and CRA both give the results that the theory predicts. Analysis of full data is always unbiased with coverage close to the $95\%$ level and the smallest standard errors of all methods. CRA is unbiased under M1 and M3 as expected,\cite{White2010b} but bias is observed under the other two missingness mechanisms. Coverage is correspondingly low when bias is present, and efficiency is lower than that in full data. 
\par
Under M1, when $x$ is MCAR, all methods appear unbiased, with comparable empirical and average model standard errors and correct coverage. This is as expected.
\par
Under M2, when $x$ is MAR conditional on $y$, CRA is severely biased in the estimate of $\beta_{0}$ and the corresponding coverage of 95\% CIs falls to 0. However, the method provides an unbiased estimate of $\beta_{x}$ with correct coverage. This result is specific to this simulation set-up, where the probability of being a complete record depends on the outcome, and the analysis model is a logistic regression. This mimics case-control sampling, where the log odds of the logistic regression is biased in case-control studies but the log odds ratio is not.\cite{White2010b, Bartlett2015} The outcome--covariate association can therefore be estimated consistently among the complete records. Standard MI and calibrated-$\delta$ adjustment MI are unbiased for both parameter estimates. Standard MI yields comparable empirical and average model standard errors and coverage attains the nominal level. In calibrated-$\delta$ adjustment MI, empirical standard errors are slightly smaller than the average model counterparts, leading to a minimal increase in coverage. 
\par
Under M3, when $x$ is MNAR dependent on $x$, CRA yields unbiased estimates of both parameters. Standard MI is biased in the estimate of $\beta_{0}$ but provides an unbiased estimate of $\beta_{x}$ due to the symmetry property of the odds ratios. Generally, in logistic regression with an incomplete covariate $x$, when the missingness mechanism is such that both standard MI and CRA are unbiased, standard MI tends not to be more efficient than CRA in estimating $\beta_{x}$.\cite{White2010b} This is because without auxiliary variables in the imputation model, standard MI does not carry any extra information on the odds ratio compared to CRA. This is seen in the simulation results for $\beta_{x}$ under models M1--M3. Under M3, calibrated-$\delta$ adjustment MI is also unbiased in both parameter estimates. Given that all three methods are unbiased for $\beta_{x}$ under M3, there is a small gain in efficiency in the estimate of $\beta_{x}$ in calibrated-$\delta$ adjustment MI, as the empirical standard error for this parameter is slightly smaller than that in CRA. Under this missingness mechanism, empirical and average model standard errors are comparable across methods; for methods that are unbiased, their corresponding coverage of $95\%$ CIs generally attains the nominal level.
\par
Under M4, when $x$ is MNAR dependent on $x$ and $y$, standard MI and CRA are again biased in both parameter estimates, leading to coverage close or equal to 0. In contrast, calibrated-$\delta$ adjustment MI produces unbiased estimates of both parameters. In this method, empirical standard errors are again slightly smaller than the average model counterparts (as seen previously under M2), which leads to coverage slightly exceeding the $95\%$ level. 
%

\subsection{When the population distribution is estimated with uncertainty}
\label{subsec3.2}
So far, the population distribution of the incomplete covariate that is used to derive the calibrated-$\delta$ adjustment is assumed to be obtained from a population census or equivalent. In other words, it is assumed that there is no uncertainty associated with estimating the reference distribution, and hence, the adjustment. In calibrated-$\delta$ adjustment MI, we believe that the extra uncertainty in estimating the calibrated-$\delta$ adjustment should be ignored when the population distribution of the incomplete covariate is assumed to be invariant, unless the reference population is not a census or equivalent. Since MI is a Bayesian procedure in which all sources of uncertainty are modelled, this explains why, if there is uncertainty about the population distribution of the incomplete covariate, this uncertainty needs to be accounted for in the derivation of the calibrated-$\delta$ adjustment across imputations.
\par
When the population distribution of the incomplete covariate is not `known' and is estimated, a natural approach for incorporating this extra uncertainty would be to draw values of the population proportions from their distribution and calculate the calibrated-$\delta$ adjustment using these draws, so that this uncertainty is reflected in the MI variance estimation. This additional step is expected to have an effect on the between-imputation variance of Rubin's variance estimator.
\par
An extension of the simulation study presented in section \ref{subsec3.1} is conducted to explore this setting.

\subsubsection{Method}
\label{subsubsec3.2.1}
This extended simulation study of a fully observed binary outcome y and a partially observed binary covariate $x$ follows the same method described in \cref{subsubsec3.1.1}, except that two variations of the population proportions of $x$ are evaluated in the imputation step of calibrated-$\delta$ adjustment MI. The reference distribution is assumed to either come from a census or equivalent (case 1), or be estimated in an external dataset of larger size (case 2) or smaller size (case 3) than the study sample.
\par
Suppose that in an external dataset of size $n^{\text{ex}}$ which comes from the same population as the study sample, the sample proportion $\hat{p}_{x}^{\text{pop}}$ provides an unbiased estimate of the population proportion $p_{x}^{\text{pop}}$. Assuming that the sampling distribution of the sample proportions is approximately normal, its standard error is given by 
\begin{equation*}
    \text{SE}\left(\hat{p}_{x}^{\text{pop}}\right) = \sqrt{\frac{\hat{p}_{x}^{\text{pop}}\left(1-\hat{p}_{x}^{\text{pop}}\right)}{n^{\text{ex}}}}.
\end{equation*}
The data generating mechanism and analysis procedures are as follows.
\begin{enumerate}
    \item For cases 2 and 3, the following two steps are performed to incorporate the sampling behaviour of $\hat{p}_{x}^{\text{pop}}$, which is estimated in an external dataset of size $n^{\text{ex}}$, into the data generating mechanism in repeated simulations.
    \begin{enumerate}[label=\alph*.]
        \item Simulate $n^{\text{ex}} = 10\,000$ (case 2) or $1\,000$ (case 3) complete values of the binary $0/1$ covariate $x$ from the model
        \begin{equation*}
            x \sim \text{Bernoulli}\left(p_{x}^{\text{pop}} = 0.7\right);
        \end{equation*}
        \item Obtain the sample proportion $\hat{p}_{x}^{\text{pop}}$ of $x$, which is an unbiased estimate of the population proportion $p_{x}^{\text{pop}}$;
    \end{enumerate}
    \item Simulate $n=5\,000$ complete values of the binary $0/1$ covariate $x$ and binary $0/1$ covariate $y$ from the models
    \begin{align}
        &x \sim \text{Bernoulli}\left(p_{x}^{\text{pop}} = 0.7\right); \nonumber \\
        &\text{logit}\left[p\left(y=1 \mid x\right)\right] = \beta_{0} + \beta_{x}x \label{eq:moi_sim2},
    \end{align}
    where $\beta_{0}$ and $\beta_{x}$ are arbitrarily set to $\text{ln}\left(0.5\right)$ and $\text{ln}\left(1.5\right)$, respectively. The same values of the $\beta$ coefficients are used throughout to make bias comparable across all simulation settings;
    \item Simulate a binary indicator of response $r$ of $x$ from each of the selection models M1--M4 (Table \ref{tab:analytic_study_selections}). Values of $1.5$ and $-1.5$ are chosen for $\alpha_{y}$ and $\alpha_{x}$ in M2 and M3, respectively. For M4, $\alpha_{y} = 1.5$ and $\alpha_{x} = -1.5$ are used. In all selection models, $\alpha_{0}$ is altered to achieve approximately $45\%$ missing $x$. For M1, $\alpha_{0}$ is calculated directly as $\text{ln}\left(\frac{0.55}{0.45}\right)$; for M2--M4, $\alpha_{0} = -0.2; 1.35$; and $0.75$ are used;
    \item For $i=1, \ldots, 5\,000$, set $x_{i}$ to missing if $r_{i} = 0$;
    \item Impute missing values in $x$ $M=10$ times using standard MI and calibrated-$\delta$ adjustment MI in turn. For cases 2 and 3, calibrated-$\delta$ adjustment MI is performed as follows.
    \begin{enumerate}[label=\alph*.]
        \item Draw a value $\tilde{p}_{x}^{\text{pop}}$ from the normal approximation $N\left(\hat{p}_{x}^{\text{pop}}, \frac{\hat{p}_{x}^{\text{pop}}\left(1-\hat{p}_{x}^{\text{pop}}\right)}{n^{\text{ex}}}\right)$, with values of $n^{\text{ex}} = 10\,000$ (case 2) and $1\,000$ (case 3). This is done by first taking a draw $\tilde{z}$ from the standard normal distribution, $z \sim N\left(0,1\right)$, followed by drawing $\tilde{p}_{x}^{\text{pop}} = \hat{p}_{x}^{\text{pop}} + \tilde{z}\sqrt{\frac{\hat{p}_{x}^{\text{pop}}\left(\hat{p}_{x}^{\text{pop}}\right)}{n^{\text{ex}}}}$;
        \item Derive the calibrated-$\delta$ adjustment and perform MI according to the algorithm set out in \cref{subsubsec3.1.1}, using $\tilde{p}_{x}^{\text{pop}}$ as the reference proportion;
    \end{enumerate}
    \item For each MI method, fit the analysis model \eqref{eq:moi_sim2} to each completed dataset and combine the results using Rubin's rules.\cite{Rubin1987, Barnard1999}
\end{enumerate}
Step 5 is designed to mimic the full Bayesian sampling process, which is always the aim in proper (or Rubin's) MI. Again, steps 1--6 are repeated $S=2\,000$ times under each of the four selection models M1--M4, so the same set of simulated independent datasets is used to compare the two MI methods under the same missingness scenario, but a different set of datasets is generated for each missingness scenario.\cite{Burton2006} The parameters of interest are $\beta_{0}$ and $\beta_{x}$ . Bias in $\hat{\beta}_{0}$ and $\hat{\beta}_{x}$, efficiency in terms of the empirical and average model standard errors, and coverage of $95\%$ CIs are calculated over $2\,000$ repetitions for each combination of simulation settings,\cite{White2010a} with analyses of full data and complete records also provided for comparison. 
\par
All simulations are performed in Stata 14 \cite{StataCorp2015b} with \texttt{mi impute logit} for standard MI, the community-contributed command \texttt{uvis logit} \cite{Royston2004} for calibrated-$\delta$ adjustment MI, and \texttt{mi estimate: logit} for fitting the analysis model to the completed datasets and combining the results using Rubin's rules \cite{Rubin1987, Barnard1999}; simulated datasets are analysed using the community-contributed command \texttt{simsum}.\cite{White2010a}

\subsubsection{Results}
\label{subsubsec3.2.2}
\begin{figure}[htbp!]
    \centering
    \caption{Extended simulation study: bias in point estimates, empirical and average model SE, and coverage of 95\% CIs under different missingness mechanism for $x$; the population distribution of $x$ is assumed to be invariant (case 1) or estimated in an external dataset of size 10\,000 (case 2) or 1\,000 (case 3).}
    \label{fig:sim_var}
    \begin{subfigure}[b]{0.65\textwidth}
        \includegraphics[width=\textwidth]{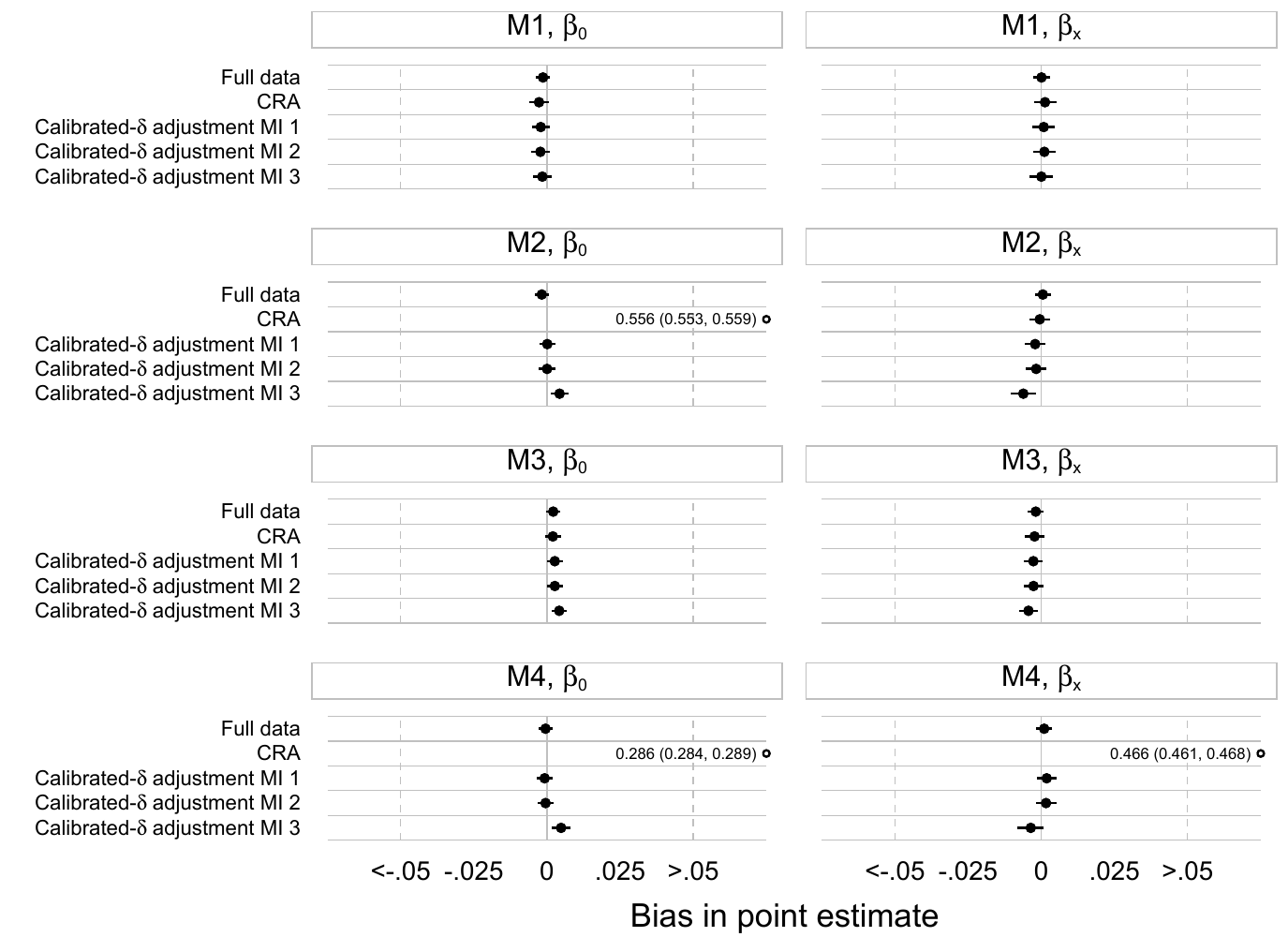}
    \end{subfigure}
    ~ 
    \begin{subfigure}[b]{0.65\textwidth}
        \includegraphics[width=\textwidth]{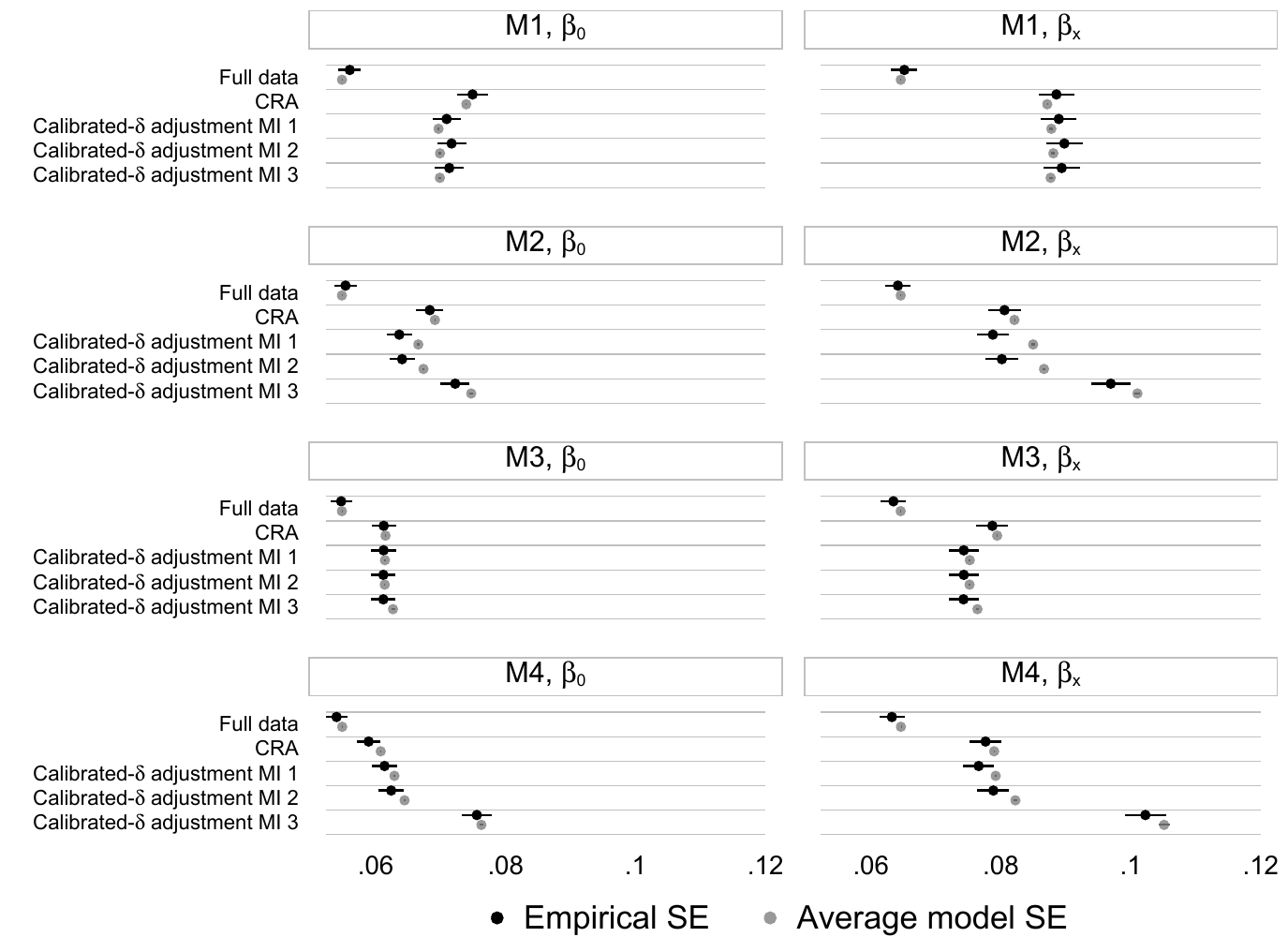}
    \end{subfigure}
    ~ 
    \begin{subfigure}[b]{0.65\textwidth}
        \includegraphics[width=\textwidth]{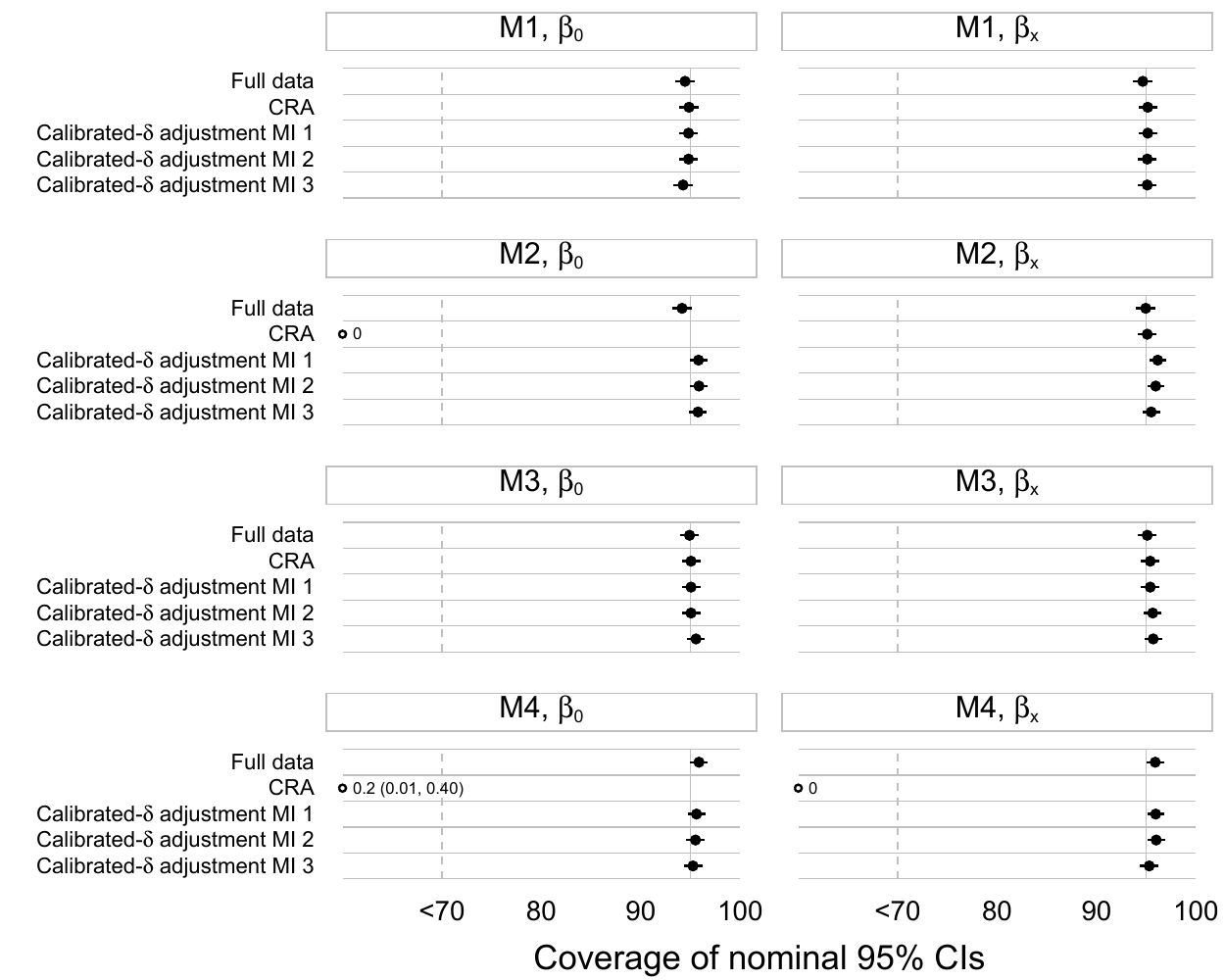}
    \end{subfigure}
\end{figure}

Results of the extended simulation study are presented in Figure \ref{fig:sim_var}.
Bias in point estimates is similar when $p_{x}^{\text{pop}}$ is invariant or estimated in a large external dataset (cases 1 and 2, respectively). Bias slightly increases, particularly under M2 and M4, when $p_{x}^{\text{pop}}$ is estimated in a small external dataset with higher variance (case 3). 
\par
Empirical and average model standard errors are comparable and remain stable for calibrated-$\delta$ adjustment MI across the three cases under M1 and M3. Under M2 and M4, the discrepancy previously seen between the empirical and average model standard errors in calibrated-$\delta$ adjustment MI (\cref{subsubsec3.1.2}) decreases in case 3 compared to cases 1 and 2. When there is increased uncertainty in estimating the population proportions of $x$ (case 3 compared to case 1), there is also a marked increase in both the empirical and average model standard errors in calibrated-$\delta$ adjustment MI. This extra uncertainty is reflected in the variation of the point estimates across the simulation repetitions according to how the simulation is set up, and is also acknowledged by an increase in the between-imputation variance component of Rubin's variance estimator (results for between-imputation variances not shown).
\par
In line with results seen for the standard errors, coverage attains the nominal level for calibrated-$\delta$ adjustment MI under M1 and M3. Under M2 and M4, since the empirical standard errors are closer to the average model standard errors in case 3 compared to case 1, the slight over-coverage of 95\% CIs seen in case 1 seems to disappear in case 3.

\section{Case study -- ethnicity and the prevalence of type 2 diabetes diagnoses in The Health Improvement Network primary care database}
\label{sec4}

This case study is conducted to illustrate the use of calibrated-$\delta$ adjustment MI for handling missing data in ethnicity in UK primary care electronic health records, when ethnicity is included as a covariate in the analysis model. In particular, this is a cross-sectional study which examines the association between ethnicity and the prevalence of type 2 diabetes diagnoses in a large UK primary care database in 2013. Prevalence of type 2 diabetes is chosen as the outcome variable to illustrate the application of the calibrated-$\delta$ adjustment MI method as developed and evaluated in \cref{sec2,sec3}.

\subsection{The Health Improvement Network database}
\label{subsec4.1}
The Health Improvement Network (THIN) \cite{IMSHealth2015} is one of the largest databases in the UK to collect information on patient demographics, disease symptoms and diagnoses, and prescribed medications in primary care. THIN contains anonymised electronic health records from over 550 general practices across the UK, with more than 12 million patients contributing data. The database is broadly generalisable to the UK population in terms of demographics and crude prevalences of major health conditions.\cite{Blak2011, Bourke2004} 
\par
Information is recorded during routine patient consultations with General Practitioners (GP) from when the patients register to general practices contributing data to THIN to when they die or transfer out. Symptoms and diagnoses of disease are recorded using Read codes, a hierarchical coding system.\cite{Chisholm1990, Dave2009} THIN also provides information on referrals made to secondary care and anonymised free text information. Patient demographics include information on year of birth, sex, and social deprivation status measured in quintiles of the Townsend deprivation score.\cite{Townsend1988} 
\par
The acceptable mortality reporting (AMR) \cite{Maguire2009} and the acceptable computer usage (ACU) \cite{Horsfall2013} dates are jointly used for data quality assurance in THIN. The AMR date is the date after which the practice is deemed to be reporting a rate of all-cause mortality sufficiently similar to that expected for a practice with the same demographics, based on data from the Office for National Statistics (ONS).\cite{Maguire2009} The ACU date is designed to exclude the transition period between the practice switching from paper-based records to complete computerisation; it is defined as the date from which the practice is consistently recording on average at least two drug prescriptions, one medical record and one additional health record per patient per year.\cite{Horsfall2013}
\par
Use of THIN for scientific research was approved by the NHS South-East Multi-Centre Research Ethics in 2003. Scientific approval to undertake this study was obtained from IQVIA World Publications Scientific Review Committee in September 2017 (SRC Reference Number: 17THIN083).

\subsection{Study sample}
\label{subsec4.2}
All individuals who are permanently registered with general practices in London contributing data to THIN are considered for inclusion in the study sample. This sample is chosen since it is not only more practical to perform MI on a smaller dataset, but also because London is the most ethnically diverse region in the UK, and hence incorrect assignment of ethnicity from imputing missing data with the White ethnic group is expected to be more apparent compared to other regions.
\par
For each individual, a start date is defined as the latest of: date of birth, ACU and AMR dates,\cite{Maguire2009, Horsfall2013} and registration date. Similarly, an end date is defined as the earliest of: date of death, date of transfer out of practice, and date of last data collection from the practice. Point prevalence of type 2 diabetes on 01 January 2013 is calculated, since THIN is a dynamic database in which individuals register with and leave their general practices at different times. Individuals are selected into the study sample if they are actively registered to THIN practices in London on 01 January 2013, and in addition they need to have been registered with the same general practices for at least 12 months by this date. This criterion is introduced to ensure that there is enough time for the individuals to have their type 2 diabetes diagnoses recorded in their electronic health data, after registration with their general practices.

\subsection{Outcome variable and main covariate}
\label{subsec4.3}
The recording of diabetes diagnoses and management in THIN is comprehensive and therefore there are several ways an individual may be identified as diabetic. For this study, an algorithm developed by Sharma et al. \cite{Sharma2016a} is used to identify individuals with diabetes mellitus, as well as to distinguish between type 1 and type 2 diabetes. According to this algorithm, individuals are identified as having diabetes if they have at least two of the following records: a diagnostic code for diabetes, supporting evidence of diabetes (e.g. screening for diabetic retinophany), or prescribed treatment for diabetes. In this study, the first record of any of these three is considered as the date of diagnosis. In addition to identifying individuals with diabetes, the algorithm also distinguishes between type 1 and type 2 diabetes based on individuals' age at diagnosis, types of treatment and timing of the diabetes diagnosis. \cite{Sharma2016a, Sharma2016b} After the study sample is selected using the method described in \cref{subsec4.2}, prevalent cases of type 2 diabetes are defined as individuals who have a diagnosis of type 2 diabetes on or before 01 January 2013.
\par
Ethnicity is typically recorded in THIN using the Read code system \cite{Chisholm1990}; it can also be recorded using free text entries. A list containing Read codes related to ethnicity is developed using a published method.\cite{Dave2009} The majority of ethnicity records are identified by searching both the medical and additional health data files for Read codes in the ethnicity code list. Minimal additional information is found by searching the pre-anonymised free text as well as other free text linked to ethnicity-related Read codes. Ethnicity is then coded into the five-level ONS classification as White, Mixed, Asian, Black, and Other ethnic groups.\cite{OfficeforNationalStatistics2012} Subsequently, the Mixed and Other ethnic groups are combined due to the small counts and heterogeneity in these two groups. Searching for ethnicity-related Read codes reveals that there is a small number of individuals with multiple inconsistent records of ethnicity. For these individuals, it can not be determined with certainty whether their ethnicity is in fact one of the recorded categories or if all the recorded categories are incorrect. Therefore, their ethnicity is set to missing for simplicity, since the issue of inconsistency in ethnicity recording is not the focus of this study.

\subsection{Statistical analysis}
\label{subsec4.4}
The analysis model in this study is a logistic regression model for a binary indicator of whether an individual has a diagnosis of type 2 diabetes on or before 01 January 2013, conditional on the individual's age in 2013, sex, Townsend deprivation score (five quintiles, from the least to the most deprived), and ethnic group (White, Asian, Black, Mixed/Other). Age is analysed in $10$-year age groups for individuals aged 0--79 years, and all individuals aged 80 years and above are grouped into the $80+$ category. Ethnicity information is extracted and categorised as described in \cref{subsec4.3}. Since this study is conducted to illustrate the application of calibrated-$\delta$ adjustment MI in a univariate missing data setting where missing data occurs in a single covariate (ethnicity), individuals with incomplete information on age, sex, and deprivation status were excluded from the analysis. 
\par
Missing values in ethnicity are handled by (i) a CRA, (ii) single imputation with the White ethnic group, (iii) standard MI, and (iv) calibrated-$\delta$ adjustment MI using the 2011 ONS census distribution of ethnicity in London \cite{OfficeforNationalStatistics2012} as the reference distribution. For MI of ethnicity, a multinomial logistic regression imputation model is constructed for ethnicity using all variables in the analysis model, including individuals' age group in 2013, sex, and quintiles of the Townsend score. In MI, the outcome variable must be explicitly included in the imputation model for the incomplete covariate. \cite{Sterne2009} Since the analysis model is a logistic regression model, the type 2 diabetes indicator is also included as a covariate in the imputation model for ethnicity. 
\par
In this study, ethnicity is analysed as a four-level categorical variable. Therefore, the calibrated-$\delta$ adjustment MI method for handling missing data in an incomplete binary covariate discussed in \cref{sec2,sec3} can be generalised for handling missing values in ethnicity as a categorical covariate. The overall proportion of the $j$th level of ethnicity, $j=1, \ldots, 4$ can be written as
\begin{equation}
    \label{eq:eth_partition}
    p\left(\text{eth} = j\right) = p\left(\text{eth} = j \mid r = 1\right) p\left(r=1\right) + p\left(\text{eth} = j \mid r = 0\right) p\left(r=0\right),
\end{equation}
where $p\left(\text{eth} = j\right)$ is available in the census; $p\left(\text{eth} = j \mid r = 1\right)$, $p\left(r=1\right)$, and $p\left(r=0\right)$ can be obtained in the observed data.
\par
A multinomial logistic regression imputation model for ethnicity conditional on age group (40--49 years old as the base level), sex (male as the base level), Townsend score (quintile 1 as the base level), and the binary indicator of type 2 diabetes (no diagnosis as the base level) is fitted to the observed data. Setting the first level of ethnicity (White, $j = 1$) as the base level to identify the model, the probability of the level $j$th of ethnicity in the observed data, $j=2, \ldots, 4$ can be written in terms of the observed-data linear predictors, $\text{linpred}_{j}^{\text{obs}}$, which is estimated from the multinomial logistic regression model for ethnicity as
\begin{equation}
    \label{eq:pethj_r1}
    p\left(\text{eth}=j \mid r=1\right) = \frac{1}{n^{\text{obs}}} \sum_{i=1}^{n^{\text{obs}}}\frac{1}{1+\sum_{j=2}^{4}\left(\text{linpred}_{ij}^{\text{obs}}\right)},
\end{equation}
where $i$ indexes individuals in the dataset, and
\begin{align}
    \label{eq:linpred_obs}
    \text{linpred}_{ij}^{\text{obs}} &= \theta_{j0}^{\text{obs}} + \sum_{a=10}^{30}  \theta_{j\text{age}_{a}}^{\text{obs}}I\left[\text{age}_{ij}=a\right] + \sum_{a=50}^{80}  \theta_{j\text{age}_{a}}^{\text{obs}}I\left[\text{age}_{ij}=a\right] + \theta_{j\text{sex}}^{\text{obs}}I\left[\text{sex}_{ij} = \text{female}\right] \nonumber \\ 
    & + \sum_{t=2}^{5}\theta_{j\text{town}_{t}}^{\text{obs}}I\left[\text{Townsend}_{ij} = t\right] + \theta_{j\text{t2d}}^{\text{obs}}I\left[\text{type 2 diabetes}_{ij} = \text{yes}\right].
\end{align}
\par
Following the methods outlined in \cref{sec3}, since covariates in the imputation model for
ethnicity are all binary or categorical, the relative risk ratios are the same among those with ethnicity observed and missing. The linear predictors in the missing data, $\text{linpred}_{j}^{\text{mis}}$, can therefore be written as
\begin{align}
    \label{eq:linpred_mis}
    \text{linpred}_{ij}^{\text{mis}} &= \left(\theta_{j0}^{\text{obs}} + \delta_{j0}\right)+ \sum_{a=10}^{30}  \theta_{j\text{age}_{a}}^{\text{obs}}I\left[\text{age}_{ij}=a\right] + \sum_{a=50}^{80}  \theta_{j\text{age}_{a}}^{\text{obs}}I\left[\text{age}_{ij}=a\right] + \theta_{j\text{sex}}^{\text{obs}}I\left[\text{sex}_{ij} = \text{female}\right] \nonumber \\ 
    & + \sum_{t=2}^{5}\theta_{j\text{town}_{t}}^{\text{obs}}I\left[\text{Townsend}_{ij} = t\right] + \theta_{j\text{t2d}}^{\text{obs}}I\left[\text{type 2 diabetes}_{ij} = \text{yes}\right],
\end{align}
where $\delta_{j0}$ is the level-$j$ intercept adjustment in the multinomial logistic regression imputation model for ethnicity. Hence, the probability of the $j$th level of ethnicity in the missing data, $j=2, \ldots, 4$, is given by
\begin{equation}
    \label{eq:pethj_r0}
    p\left(\text{eth}=j \mid r=0\right) = \frac{1}{n^{\text{mis}}} \sum_{i=1}^{n^{\text{mis}}}\frac{1}{1+\sum_{j=2}^{4}\left(\text{linpred}_{ij}^{\text{mis}}\right)}.
\end{equation}
\par
From \cref{eq:eth_partition,eq:pethj_r1,eq:linpred_obs,eq:pethj_r0,eq:linpred_mis}, to implement calibrated-$\delta$ adjustment MI, we need to find the solutions $\delta_{j0}$, $j=2, \ldots, 4$, of a system of three non-linear equations for the three categories of ethnicity. The solutions of this system of equations can be obtained simultaneously using the Stata base command \texttt{nl} \cite{StataCorp2015b} and defining a function evaluator program. Once the values of the calibrated-$\delta$ adjustments are obtained, the imputation is performed using the same procedure as outlined in \cref{subsec3.1}.
\par
Both MI methods are performed using $M=30$ imputations, and Rubin's rules \cite{Rubin1987, Barnard1999} are used to obtain estimates of association and standard errors. All analyses are conducted using Stata 14, \cite{StataCorp2015b} where \texttt{mi impute mlogit} is used for standard MI, the community-contributed command \texttt{uvis mlogit} \cite{Royston2004} for calibrated-$\delta$ adjustment MI, and \texttt{mi estimate: logit} for performing the main analysis in the completed datasets and obtaining the final results using Rubin's rules.\cite{Rubin1987, Barnard1999}

\subsection{Results}
\label{subsec4.5}
\begin{figure}[ht!]
    \centering
    \caption{Case study: flowchart of selection criteria for THIN sample.}
    \label{fig:flowchart}
    \includegraphics[scale=0.35]{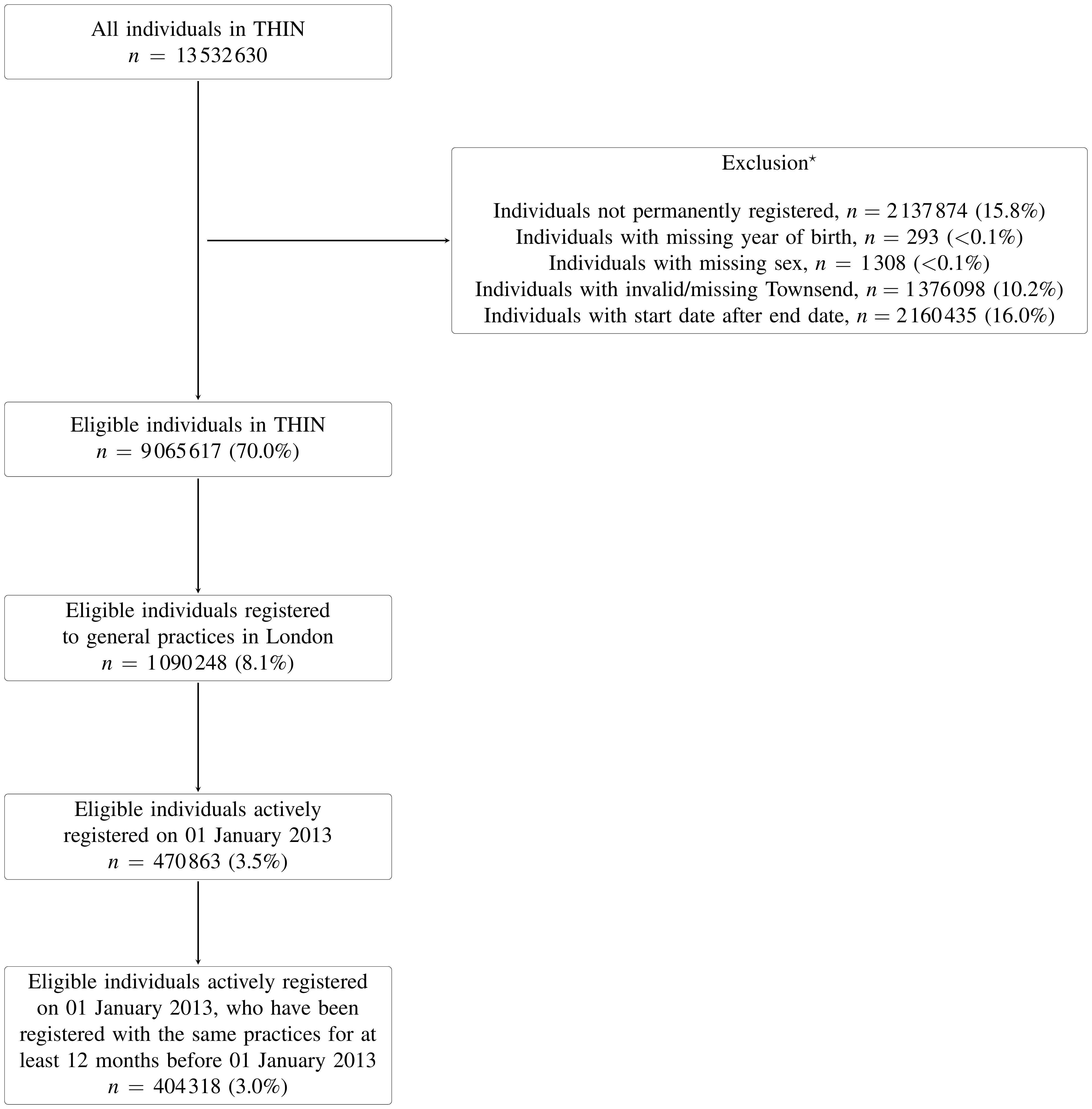}
    \caption*{\textsuperscript{$\star$}An individual can be excluded from the study sample due to more than one criterion.}
\end{figure}

Figure \ref{fig:flowchart} depicts a flowchart of the selection criteria used to obtain the relevant sample for this study. In total, data from 13\,532\,630 individuals are extracted from THIN, of which 2\,137\,874 (15.8\%) individuals are not permanently registered, 293 (less than 0.1\%) individuals do not have their year of birth recorded, 1\,308 (less than 0.1\%) individuals have missing sex, 1\,376\,098 (10.2\%) individuals have invalid or missing Townsend score, and 2\,160\,435 (16.0\%) have their start date after their end date. Applying the selection criteria results in 9\,065\,617 (70.0\%) individuals who are eligible for inclusion in this study. In this eligible sample, there are 1\,090\,248 (8.1\%) individuals who are registered to THIN general practices in London, of whom 470\,863 (3.5\%) individuals are actively registered on 01 January 2013. Finally, $n=404\,318 \left(3.0\%\right)$ individuals have at least 12 months of follow-up by 01 January 2013 and make up the sample for this study.
Table \ref{tab:example2_vars} presents a summary of variables considered in this study. The sample comprises $51\%$ women; the majority of individuals in the sample (approximately $80\%$) are below 60 years of age; there are slightly more than $70\%$ of the individuals with quintiles of the Townsend score of 3 and above; and $5.5\%$ of the individuals have a diagnosis of type 2 diabetes on or before 01 January 2013.
\par
\begin{figure}[t!]
    \centering
    \caption{Case study: distribution of four-level ethnicity in different methods for handling
missing ethnicity data, compared to the 2011 ONS census distribution for London (horizontal black lines).}
    \label{fig:dist_eth}
    \includegraphics[scale=0.8]{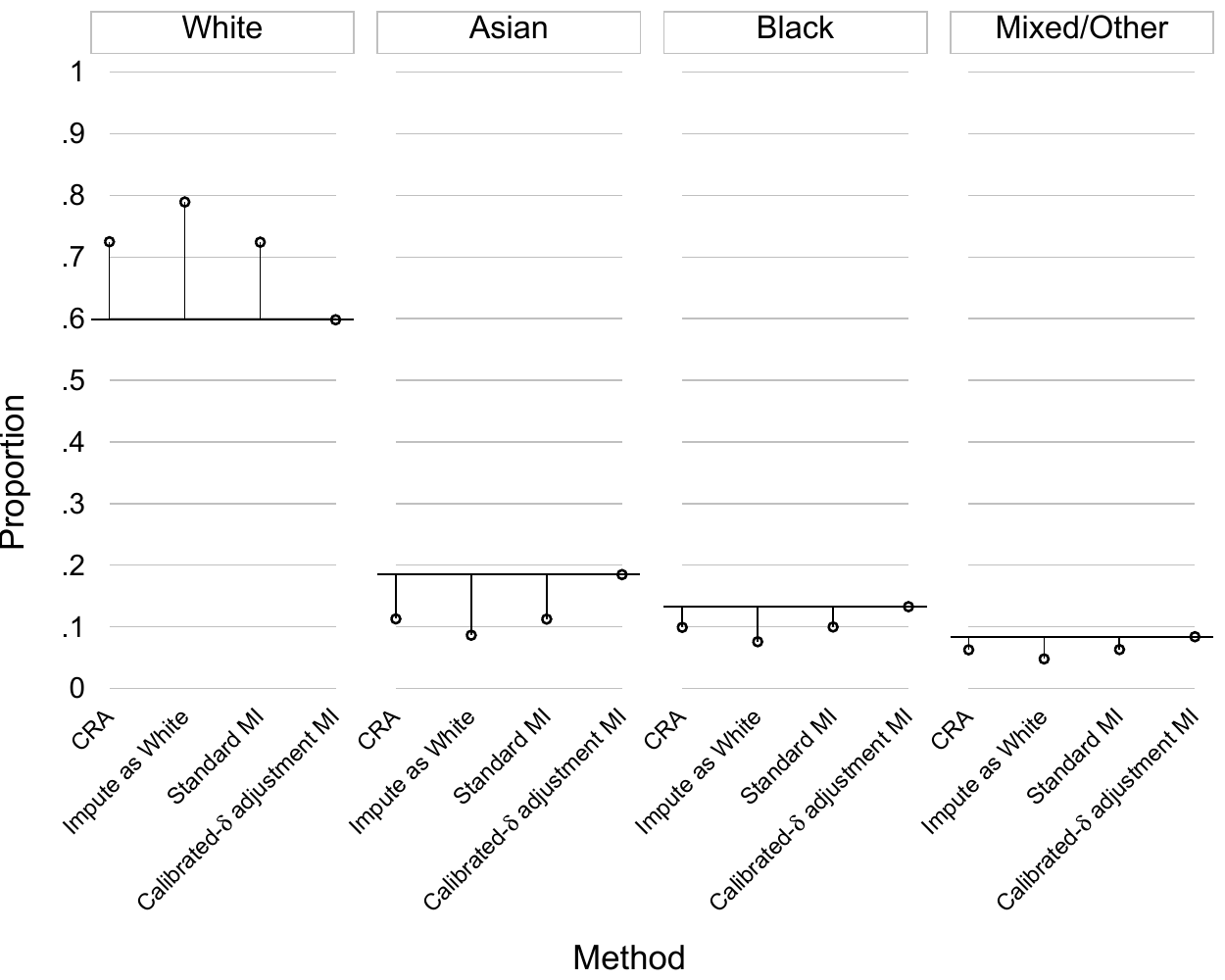}
\end{figure}

Ethnicity is recorded for $309\,684 \left(76.6\%\right)$ and missing for $94\,634 \left(23.4\%\right)$ individuals (Table \ref{tab:dist_etht2d}). Among individuals with ethnicity recorded, the estimated proportion of the White ethnic group is higher, and the non-White ethnic groups lower compared to the corresponding ethnic breakdown in the 2011 ONS census data for London (Table \ref{tab:dist_etht2d}). Single imputation with the White ethnic group further overestimates the White group and underestimates the other non-White groups, under the assumption that the ethnicity distribution in THIN should match that in the census (Table \ref{tab:dist_etht2d}).
\begin{table}[ht!]
	\renewcommand{\arraystretch}{1.1}
	\setlength{\tabcolsep}{5pt}
	\centering
	\caption{Case study: summary of variables in the analysis; $n = 404\,318$.}
	\begin{subtable}{\linewidth}
	\centering
	\subcaption{Distribution of age group, sex, Townsend deprivation score, and type 2 diabetes diagnoses.}	
	\label{tab:example2_vars}
	\begin{tabular}{lcc}
	\toprule
Variable             & Frequency & \% \\ \midrule
\textit{Age group (years)}            &           &            \\
0--9 &  41\,601       & 10.29 \\
10--19                & 45\,664   & 11.29      \\
20--29                & 50\,065   & 12.38      \\
30--39                & 65\,695   & 16.25      \\
40--49                & 64\,837   & 16.04      \\
50--59                & 53\,272   & 13.18      \\
60--69                & 39\,427   & 9.75       \\
70--79                & 25\,348   & 6.27       \\
80+                  & 18\,409   & 4.55       \\
\textit{Sex} & & \\
Male & 198\,301       & 49.05 \\
Female          & 206\,017  & 50.95      \\
\textit{Townsend score}       &           &            \\
Quintile 1 (least deprived) & 48\,934       & 12.10 \\
Quintile 2           & 64\,788   & 16.02      \\
Quintile 3           & 101\,305  & 25.06      \\
Quintile 4           & 102\,626  & 25.38      \\
Quintile 5 (most deprived)          & 86\,665   & 21.43      \\
\textit{Type 2 diabetes} & 22\,100   & 5.47      \\ \bottomrule
	\end{tabular}
    \end{subtable}
    \vskip 15pt
    \begin{subtable}{\linewidth}
	\centering
	\subcaption{Distribution of ethnicity when missing values are included, excluded, and imputed with the White ethnic group.}
	\label{tab:dist_etht2d}
	\begin{tabular}{lcccccc}
\toprule
Ethnicity         & Frequency & \begin{tabular}[c]{@{}c@{}}\%\\ including \\ missing\end{tabular} & \begin{tabular}[c]{@{}c@{}}\%\\ excluding \\ missing\end{tabular} & \begin{tabular}[c]{@{}c@{}}Frequency\\ missing\\ imputed\\ with White\end{tabular} & \begin{tabular}[c]{@{}c@{}}\%\\ missing\\ imputed\\ with White\end{tabular} & \begin{tabular}[c]{@{}c@{}}\%\\ 2011 ONS \\ census \\ London\end{tabular} \\ \midrule
White             & 224\,403  & 55.50                                                               & 72.46                                                              & 319\,037                                                                           & 78.91                                                                               & 59.8                                                                            \\
Asian             & 35\,027   & 8.66                                                               & 11.31                                                              & 35\,027                                                                            & 8.66                                                                               & 18.8                                                                            \\
Black             & 30\,771   & 7.61                                                               & 9.94                                                               & 30\,771                                                                            & 7.61                                                                                & 13.3                                                                            \\
Other             & 19\,483   & 4.82                                                               & 6.29                                                               & 19\,483                                                                            & 4.82                                                                                & 8.4                                                                             \\
Missing           & 94\,634   & 23.41                                                              &                                                                    &                                                                                    &                                                                                     &                                                                                 \\\midrule
$\sum$ including missing & 404\,318  &                                                                    &                                                                    &                                                                                    &                                                                                     &                                                                                 \\
$\sum$ excluding missing & 309\,684  &                   \\ \bottomrule      
	\end{tabular}
    \end{subtable}
\end{table}
\clearpage
Figure \ref{fig:dist_eth} shows the distribution of four-level ethnicity after missing values in ethnicity are handled by the various methods for missing data. CRA, single imputation with the White ethnic group, and standard MI overestimate the White group while underestimating the other non-White ethnic proportions, compared to the corresponding census statistics. In calibrated-$\delta$ adjustment MI, the majority of missing values in ethnicity are imputed with the Asian and Black groups. This method recovers the ethnic breakdown in the census as expected, since the census distribution is used as the reference.
\par
\begin{figure}[t!]
    \centering
    \caption{Case study: estimated odds ratio of type 2 diabetes diagnosis for age group (base level: 40-49 years), sex (base level: male), social deprivation status (base level: quintile 1 of the Townsend score), and ethnicity (base level: White) in different methods for handling missing ethnicity data.}
    \label{fig:combine_methodf}
    \includegraphics[scale = 1.27]{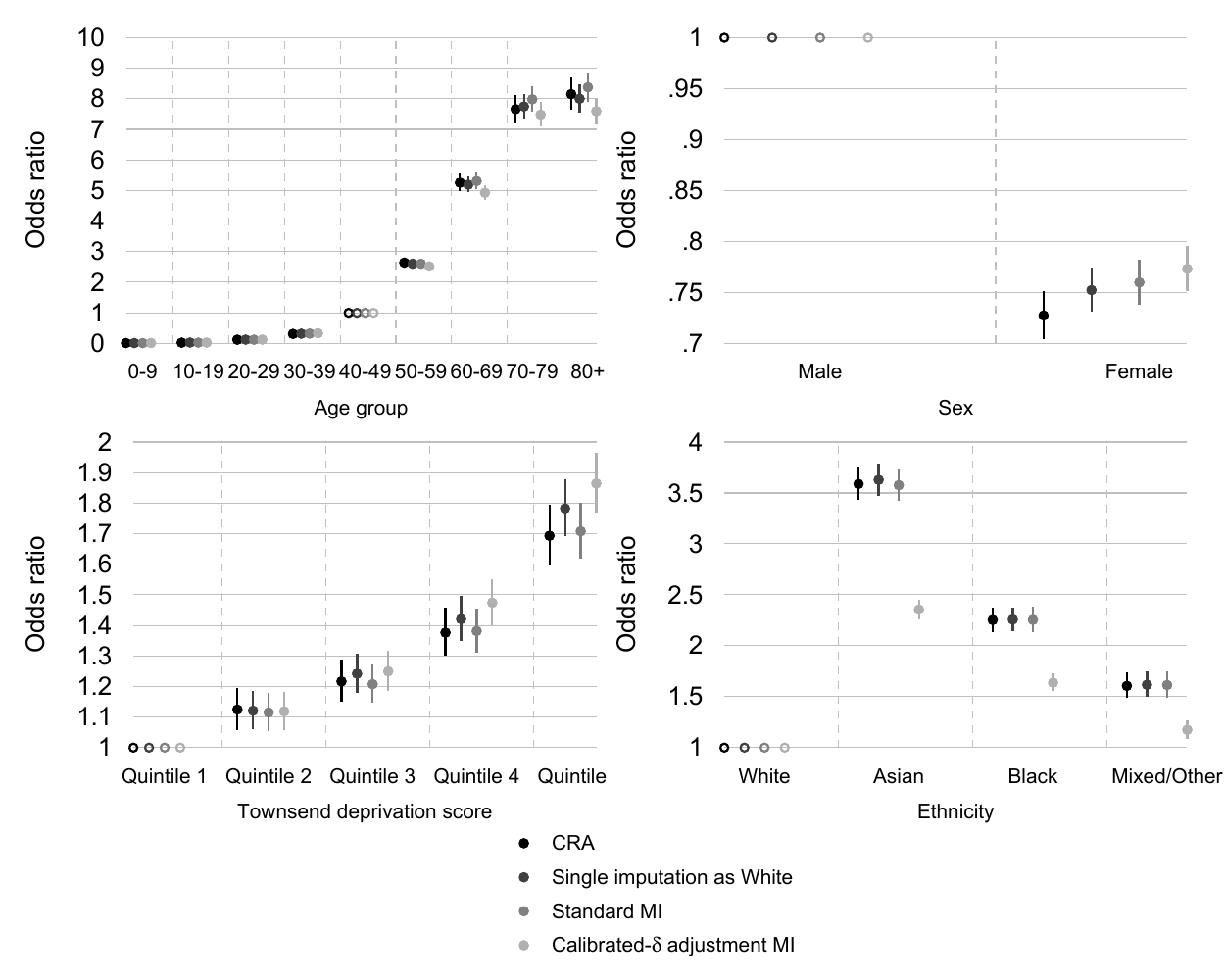}
\end{figure}

Figure \ref{fig:combine_methodf} and Table \ref{tab:case_study_or} present estimated odds ratios of type 2 diabetes diagnosis and $95\%$ CIs for age group, sex, Townsend score, and ethnicity in the analysis model. Age 40--49 years, male, quintile 1, and the White ethnic group are selected as base levels for age group, sex, Townsend score, and ethnicity, respectively. $M = 30$ imputations produce Monte Carlo errors for point estimates of less than 10\% of the estimated standard errors for all parameters. The relative efficiency versus an infinite number of imputations is above 0.988 for all parameter estimates and MI methods. Overall, the odds of being diagnosed with type 2 diabetes increase relatively smoothly with older age groups and higher quintiles of the Townsend score; are lower in women compared to men; and are higher in the Asian, Black, and Mixed/Other ethnic groups compared to the White group in all methods for handling missing data in ethnicity.
\par
Compared to the other three methods under consideration, calibrated-$\delta$ adjustment MI produces comparable estimated odds ratios for the younger age groups, and smaller estimated odds ratios for the older ($60+$) age groups. Calibrated-$\delta$ adjustment MI leads to slightly higher estimated odds ratio for women compared to CRA, single imputation with the White ethnic group, and standard MI; this increase is towards the null. All missing data methods produce odds ratios that increase with more deprived quintiles of the Townsend score. Calibrated-$\delta$ adjustment MI yields similar estimated odds ratios compared to the other methods for the first three quintiles of the Townsend score, and higher estimates for the top two quintiles.
\par
The most noticeable differences in point estimates associated with the prevalence of type 2 diabetes diagnoses are seen in the estimated odds ratios for ethnicity. CRA, single imputation, and standard MI again return similar results, in which the odds of having a diagnosis of type 2 diabetes are around 3.6 times higher in the Asian ethnic group compared to the White group, and individuals in the Black ethnic group are about 2.3 times more likely to receive a diagnosis of type 2 diabetes compared to those of White ethnic background. Single imputation with the White ethnic group slightly increases the estimated odds ratios for the non-White groups. This is because explanatory analyses conducted to examine predictors of both ethnicity and missingness in ethnicity suggest that individuals with missing ethnicity are, on average, less likely to have a diagnosis of type 2 diabetes (OR of observing ethnicity for type 2 diabetes (adjusted for age group, sex, Townsend score) = 1.39, $95\%$ CI 1.34 to 1.44, full results not shown). Replacing missing values with the White ethnic group means that this group will contain a lower percentage of type 2 diabetes diagnoses, which implies that the estimated odds ratios for the non-White ethnic groups will increase. Compared to CRA, single imputation with the White ethnic group, and standard MI, calibrated-$\delta$ adjustment MI leads to a reduction in the estimated odds ratios for the non-White ethnic groups (Figure \ref{fig:combine_methodf} and Table \ref{tab:case_study_or}). For these groups, the $95\%$ CIs of the ethnicity point estimates in calibrated-$\delta$ adjustment MI do not cross that of the other methods.
\par
Fraction of missing information (FMI) \cite{White2011} for the estimates of association between ethnicity and the prevalence of type 2 diabetes diagnoses was 0.132 (Monte Carlo standard error (MCSE) $=0.003$); 0.193 (MCSE $= 0.05$); 0.230 (MCSE $= 0.066$) for Asian, Black, and Mixed/Other ethnic group, respectively in standard MI. The corresponding quantities for these three groups in calibrated-$\delta$ adjustment MI are 0.283 (MCSE $=0.052$); 0.245 (MCSE $=0.045$); 0.327 (MCSE $=0.051$). Calibrated-$\delta$ adjustment MI appears to have higher FMI compared to standard MI. This could be explained by the fact that non-White ethnic groups, which are under-represented in the observed data, are imputed more often in calibrated-$\delta$ adjustment MI than in standard MI. Therefore, the between-imputation variance relies on more imputed values in the non-White ethnic groups and less frequently imputed values in the White group, which leads to the non-White proportion estimates being more variable across the completed datasets. 
\begin{landscape}
\begin{table}
\centering
\caption{Case study: adjusted ORs and $95\%$ CIs from a multivariable logistic regression model for the prevalence of type 2 diabetes diagnoses, conditional on age group in 2013, sex, Townsend deprivation score, and ethnic group in different methods for handling missing data in ethnicity, $n=404\,138$.}
\label{tab:case_study_or}
\begin{tabular}{lcccccccc}
\toprule
               & \multicolumn{2}{c}{CRA} & \multicolumn{2}{c}{Single imputation with White} & \multicolumn{2}{c}{Standard MI} & \multicolumn{2}{c}{Calibrated-$\delta$ adjustment MI} \\ \cmidrule(l{2pt}r{2pt}){2-3} \cmidrule(l{2pt}r{2pt}){4-5} \cmidrule(l{2pt}r{2pt}){6-7} \cmidrule(l{2pt}r{2pt}){8-9}
               & OR     & 95\% CI        & OR                 & 95\% CI                     & OR         & 95\% CI            & OR                    & 95\% CI                       \\ \midrule 
\textit{Age group (years)}      &        &                &                    &                             &            &                    &                       &                               \\
0-9            & 0.010  & 0.006 to 0.016 & 0.010              & 0.006 to 0.016              & 0.010      & 0.006 to 0.016     & 0.010                 & 0.006 to 0.017                \\
10-19          & 0.022  & 0.016 to 0.032 & 0.026              & 0.020 to 0.035              & 0.025      & 0.019 to 0.033     & 0.025                 & 0.019 to 0.033                \\
20-29          & 0.120  & 0.103 to 0.139 & 0.122              & 0.107 to 0.139              & 0.120      & 0.106 to 0.137     & 0.122                 & 0.107 to 0.139                \\
30-39          & 0.308  & 0.283 to 0.336 & 0.316              & 0.292 to 0.342              & 0.320      & 0.296 to 0.347     & 0.330                 & 0.305 to 0.357                \\
40-49          & 1      &                & 1                  &                             & 1          &                    &                       &                               \\
50-59          & 2.641  & 2.495 to 2.796 & 2.605              & 2.474 to 2.743              & 2.604      & 2.473 to 2.742     & 2.516                 & 2.390 to 2.649                \\
60-69          & 5.255  & 4.968 to 5.559 & 5.190              & 4.933 to 5.46               & 5.309      & 5.044 to 5.587     & 4.928                 & 4.685 to 5.184                \\
70-79          & 7.662  & 7.230 to 8.120 & 7.748              & 7.352 to 8.166              & 7.984      & 7.573 to 8.417     & 7.484                 & 7.102 to 7.886                \\
80+            & 8.154  & 7.655 to 8.685 & 8.003              & 7.560 to 8.472              & 8.379      & 7.910 to 8.876     & 7.596                 & 7.175 to 8.043                \\
\textit{Sex}            &        &                &                    &                             &            &                    &                       &                               \\
Male           & 1      &                & 1                  &                             & 1          &                    & 1                     &                               \\
Female         & 0.727  & 0.704 to 0.751 & 0.752              & 0.731 to 0.774              & 0.760      & 0.738 to 0.782     & 0.773                 & 0.751 to 0.796                \\
\textit{Townsend score} &        &                &                    &                             &            &                    &                       &                               \\
Quintile 1 (least deprived)     & 1      &                & 1                  &                             & 1          &                    & 1                     &                               \\
Quintile 2     & 1.125  & 1.057 to 1.196 & 1.121              & 1.060 to 1.185              & 1.115      & 1.054 to 1.179     & 1.119                 & 1.058 to 1.183                \\
Quintile 3     & 1.217  & 1.149 to 1.288 & 1.242              & 1.180 to 1.307              & 1.208      & 1.147 to 1.272     & 1.249                 & 1.187 to 1.316                \\
Quintile 4     & 1.376  & 1.300 to 1.457 & 1.420              & 1.349 to 1.496              & 1.381      & 1.312 to 1.455     & 1.474                 & 1.400 to 1.553                \\
Quintile 5 (most deprived)    & 1.693  & 1.596 to 1.796 & 1.783              & 1.691 to 1.879              & 1.708      & 1.619 to 1.802     & 1.864                 & 1.768 to 1.966                \\
\textit{Ethnic group}   &        &                &                    &                             &            &                    &                       &                               \\
White          & 1      &                & 1                  &                             & 1          &                    & 1                     &                               \\
Asian          & 3.588  & 3.431 to 3.753 & 3.629              & 3.474 to 3.789              & 3.577      & 3.425 to 3.735     & 2.355                 & 2.259 to 2.456                \\
Black          & 2.253  & 2.135 to 2.378 & 2.257              & 2.142 to 2.379              & 2.254      & 2.136 to 2.379     & 1.638                 & 1.555 to 1.725                \\
Mixed/Other    & 1.606  & 1.486 to 1.736 & 1.617              & 1.497 to 1.746              & 1.615      & 1.491 to 1.749     & 1.174                 & 1.085 to 1.270                \\ \bottomrule
\end{tabular}
\end{table}
\end{landscape}
%
\section{Discussion}
\label{sec5}
%
%
Our proposed calibrated-$\delta$ adjustment MI method for missing data in a binary/categorical covariate involves utilising population-level information about the incomplete covariate to generate a calibrated-$\delta$ adjustment, which is then used in the intercept of the imputation model in order to improve the analysis of data suspected to be MNAR. The development of this method was motivated by van Buuren et al.'s \cite{VanBuuren1999} $\delta$ (offset) approach in MI, but where $\delta$ is derived based on external information instead of chosen arbitrarily or based on expert's belief (which is arguably not arbitrary, but can be subjective). Direct linkage to external data has also increasingly been used for the analysis of missing data generated by a MNAR mechanism.\cite{Cornish2015} However, external linked data might not always be available, or the linkage might not be possible, whereas our proposed calibrated-$\delta$ adjustment MI method does not require records from the same individuals to be directly linked between the datasets. 
\par
Under the MNAR assumption of missing data, MI results rely on subtle, untestable assumptions, and may depend heavily on the particular way the missingness mechanism is modelled. This issue emphasises the central role of sensitivity analysis, which explores how inference may vary under different missingness mechanisms.\cite{Kenward2007} MI offers flexibility for sensitivity analysis, since the imputation model can be tuned to incorporate possible departures from the MAR assumption.\cite{Kenward2007, White2011} Unfortunately, a sensitivity analysis is often not performed or reported sufficiently in practice,\cite{Wood2004, HayatiRezvan2015} a tendency abetted by the practical constraints of many applied projects. When the population-level information about the incomplete covariate is available, our proposed calibrated-$\delta$ adjustment MI method provides a useful tool for performing a single, calibrated sensitivity analysis to assess the impact of potential departures from the MAR assumption. 
\par
The analytic study of a $2\times2$ contingency table with a binary outcome variable $y$ and a binary covariate $x$ gave insights into how the method works, and will work for more general contingency table settings with one incomplete variable. The analytic study explored the appropriate derivation of the calibrated-$\delta$ adjustment under increasingly complex missingness mechanisms. We showed that when data in $x$ were MNAR dependent on $x$ or both $x$ and $y$, appropriately adjusting the intercept of the imputation model sufficiently corrected bias in the analysis model's parameter estimates. Based on this setting, simulation studies were conducted to explore scenarios when the population distribution of $x$ was either invariant (i.e. `known') or estimated in an external dataset with uncertainty. Calibrated-$\delta$ adjustment MI was shown to perform as well as standard MI in terms of bias when data were MAR. Further, calibrated-$\delta$ adjustment MI also produced unbiased parameter estimates with good coverage, and was preferred to standard MI under the two general MNAR mechanisms being evaluated.
\par
In the analytic and simulation studies, we did not consider the MNAR selection model where the probability of observing $x$ depends on both $x$, $y$, and their interaction. We suspect that calibrated-$\delta$ adjustment MI with a single intercept adjustment calculated based on the marginal distribution of $x$ alone will not fully correct bias introduced by this missingness mechanism; and that an additional sensitivity parameter for the $x$--$y$ association is present. Information about the population distribution of $x$ conditional on $y$ might be required to produce unbiased estimates when the probability of observing $x$ given $x$ differs across the levels of $y$. However, such information might not always be available in practice. Similarly, when the outcome variable $y$ is continuous, a second sensitivity parameter for the covariate--outcome association in the imputation model is needed; we will explore this setting in another paper.
\par
In the case study which examined the association between ethnicity and the prevalence of type 2 diabetes diagnoses in THIN, calibrated-$\delta$ adjustment MI using information from census data yielded a more plausible estimate of the ethnicity distribution compared to CRA, single imputation of missing values with the White ethnic group, and standard MI. Subsequently, estimates of association for the non-White ethnic groups produced by calibrated-$\delta$ adjustment MI were lower than that in the other methods. Previously, it was found that ethnicity was more likely to be recorded for individuals with a diagnosis of type 2 diabetes. By imputing missing values with the non-White ethnic groups more frequently, calibrated-$\delta$ adjustment MI led to a decrease in the percentage of prevalent type 2 diabetes cases among these groups, which we thought was the primary reason explaining the lower odds ratios compared to the other methods. In addition, it was also possible that the explanatory power of ethnicity for type 2 diabetes was partially diluted by the stronger effect of deprivation status, which compensated for the reduction in the odds ratios for ethnicity. The odds ratios for Townsend deprivation score were higher in calibrated-$\delta$ adjustment MI compared to CRA for the top two quintiles. These findings seemed to suggest that some effect of ethnicity was absorbed in Townsend score in calibrated-$\delta$ adjustment MI, where deprivation status explained some of the effect which might otherwise have been explained by ethnicity. This could be attributed to a possibility that individuals of Asian and Black ethnic groups, whose ethnicity was not recorded, were more likely to belong to the more deprived quintiles of the Townsend score. 
%
%
%
\par
Given the missingness mechanisms considered thus far for the development of calibrated-$\delta$ adjustment MI in \cref{sec2,sec3}, results in the case study suggested a potential departure from the MAR assumption for missingness in ethnicity. This was because, conditional on the outcome variable type 2 diabetes and other fully observed variables included in the analysis model, standard MI did not yield a distribution of ethnicity that was comparable to the census ethnic breakdown. Ethnicity was also not likely to be MNAR dependent only on the values of ethnicity, since the point estimates in CRA and standard MI were broadly comparable. Results from the exploratory analyses examining the associations between covariates in the imputation model for ethnicity and missingness in ethnicity among the complete records suggested that age group, sex, Townsend score, and type 2 diabetes were factors likely to be associated with whether ethnicity was recorded. This finding indicated that ethnicity was likely to be MNAR depending on the ethnic groups, fully observed outcome variable (type 2 diabetes diagnoses), as well as other fully observed covariates in the analysis model (age group, sex, and deprivation status).
\par
The major strength of calibrated-$\delta$ adjustment MI is its flexibility to be adapted to impute variables in a given dataset whose distributions might be available in some external data. Here we used census data for ethnicity in primary care electronic health records, but information obtained from other nationally representative datasets (such as the Health Survey for England \cite{UKDataService}) could similarly be used to impute missing data in other health indicators routinely recorded in primary care such as smoking status or alcohol consumption. In such instances, the variability associated with estimating the reference distribution used for calibration needs to be accounted for in calibrated-$\delta$ adjustment MI as illustrated in \cref{subsec3.2}, although this source of uncertainty might be negligible depending on the size of the external dataset.
\par
Throughout this paper, we restricted our development of calibrated-$\delta$ adjustment MI to the case of a single partially observed covariate. However, we believe this approach can be extended for handling missing data in more than one variable. Multivariate imputation by chained equations (MICE) \cite{VanBuuren1999, VanBuuren2011} is a popular procedure for performing MI of multivariate missing data, and is commonly implemented under the MAR assumption.\cite{Marston2010, Marston2014} MICE is an iterative procedure which requires the specification of an imputation model for each incomplete variable, conditional on all other variables. Our proposed univariate calibrated-$\delta$ adjustment MI method can, in principle, be embedded into MICE to impute certain MNAR variables whose distributions are available externally, while the standard MI method can be used for the imputation of other variables assuming data are MAR. Under the MICE framework, when there are several MNAR variables to be imputed, information from more than one external data source can potentially be drawn on and utilised in calibrated-$\delta$ adjustment MI for these variables. 
\par
Finally, returning to the analytic and simulation studies, we did not consider the setting where both the outcome variable $y$ and the covariate $x$ are incomplete. When $y$ is MNAR dependent on its values and in addition to the population information on $x$ we can obtain the marginal distribution of $y$ from an external dataset, then this information can be used in calibrated-$\delta$ adjustment MI for $y$ when $y$ is imputed in the MICE algorithm. If $y$ is MAR then there must be some artificial mechanism whereby the dataset is divided into two subsets; one where $y$ is MAR dependent on the observed values of $x$ and another one where $x$ is MNAR dependent on its values. In this setting, our proposed MI method should work for $x$ when it is imputed in the MICE algorithm. The more complex missingness settings involving several incomplete covariates are subjected to on-going work and will be reported in the future. 

\section*{Acknowledgements}
Tra My Pham was supported by awards to establish the Farr Institute of Health Informatics Research, London, from the Medical Research Council, Arthritis Research UK, British Heart Foundation, Cancer Research UK, Chief Scientist Office, Economic and Social Research Council, Engineering and Physical Sciences Research Council, NIHR, National Institute for Social Care and Health Research, and Wellcome Trust (grant MR/K006584/1); and the NIHR School for Primary Care Research (project number 379). James Carpenter and Tim Morris were supported by the Medical Research Council (grant numbers MC\_UU\_12023/21 and MC\_UU\_12023/29).

\section*{Conflict of interest}
The authors declare no potential conflict of interests.

\bibliography{ama}%

\begin{thebibliography}{10}
\providecommand{\url}[1]{\texttt{#1}}
\providecommand{\urlprefix}{Available from: }
\expandafter\ifx\csname urlstyle\endcsname\relax
  \providecommand{\doi}[1]{doi:\discretionary{}{}{}#1}\else
  \providecommand{\doi}{doi:\discretionary{}{}{}\begingroup
  \urlstyle{rm}\Url}\fi

\bibitem{Rubin1987}
Rubin DB. {\it {Multiple imputation for nonresponse in surveys}}.
\newblock New York: Wiley; 1987.

\bibitem{Sterne2009}
Sterne JAC, White IR, Carlin JB, et al. {Multiple imputation for missing data
  in epidemiological and clinical research: potential and pitfalls}.  {\it BMJ.
  }2009;338:b2393.

\bibitem{Klebanoff2008}
Klebanoff MA, Cole SR. {Use of multiple imputation in the epidemiologic
  literature}.  {\it American Journal of Epidemiology. }2008;168(4):355--357.

\bibitem{StataCorp2015a}
StataCorp. {\it {Stata multiple-imputation reference manual: release 14}}.
\newblock StataCorp LP; 2015.

\bibitem{VanBuuren2011}
van Buuren S. {Multiple imputation of discrete and continuous data by fully
  conditional specification}.  {\it Statistical Methods in Medical Research.
  }2007;16(3):219--242.

\bibitem{Yuan2011}
Yuan Y. {Multiple Imputation Using SAS Software}.  {\it Journal of Statistical
  Software. }2011;45(6):1--25.

\bibitem{Barnard1999}
Barnard J, Rubin DB. {Small-sample degrees of freedom with multiple
  imputation}.  {\it Biometrika. }1999;86(4):948--955.

\bibitem{Little-Rubin2002}
Little RJA, Rubin DB. {\it {Statistical analysis with missing data}}.
\newblock New Jersey: John Wiley {\&} Sons, Inc.; 2002.

\bibitem{Little1993}
Little RJA. {Pattern-mixture models for multivariate incomplete data}.  {\it
  Journal of the American Statistical Association. }1993;88(421):125--134.

\bibitem{Little1994}
Little RJA. {A class of pattern-mixture models for multivariate incomplete
  data}.  {\it Biometrika. }1994;81(3):471--483.

\bibitem{White2011}
White IR, Royston P, Wood AM. {Multiple imputation using chained equations:
  issues and guidance for practice}.  {\it Statistics in Medicine.
  }2011;30:377--399.

\bibitem{Collins2001}
Collins LM, Schafer JL, Kam CM. {A comparison of inclusive and restrictive
  strategies in modern missing data procedures}.  {\it Psychological Methods.
  }2001;6(4):330--351.

\bibitem{VanBuuren1999}
van Buuren S, Boshuizen HC, Knook DL. {Multiple imputation of missing blood
  pressure covariates in survival analysis}.  {\it Statistic in Medicine.
  }1999;18:681--694.

\bibitem{Kumarapeli2006}
Kumarapeli P, Stepaniuk R, {De Lusignan} S, Williams R, Rowlands G. {Ethnicity
  recording in general practice computer systems}.  {\it Journal of Public
  Health. }2006;28(3):283--287.

\bibitem{Aspinall2007}
Aspinall PJ, Jacobson B. {Why poor quality of ethnicity data should not
  preclude its use for identifying disparities in health and healthcare}.  {\it
  Quality and Safety in Health Care. }2007;16:176--180.

\bibitem{Mathur2013b}
Mathur R, Bhaskaran K, Chaturvedi N, et al. {Completeness and usability of
  ethnicity data in UK-based primary care and hospital databases}.  {\it
  Journal of public health (Oxford, England). }2014;36(4):684--692.

\bibitem{Osborn2015}
Osborn DP, Hardoon S, Omar RZ, et al. {Cardiovascular risk prediction models
  for people with severe mental illness: results from the prediction and
  management of cardiovascular risk in people with severe mental illnesses
  (PRIMROSE) research program}.  {\it JAMA Psychiatry. }2015;72(2):143--151.

\bibitem{Hippisley-Cox2008}
Hippisley-Cox J, Coupland C, Vinogradova Y, et al. {Predicting cardiovascular
  risk in England and Wales: prospective derivation and validation of QRISK2}.
  {\it BMJ. }2008;336:a332.

\bibitem{Marston2010}
Marston L, Carpenter JR, Walters KR, Morris RW, Nazareth I, Petersen I. {Issues
  in multiple imputation of missing data for large general practice clinical
  databases}.  {\it Pharmacoepidemiology and Drug Safety. }2010;19:618--626.

\bibitem{Marston2014}
Marston L, Carpenter JR, Walters KR, et al. {Smoker, ex-smoker or non-smoker?
  The validity of routinely recorded smoking status in UK primary care: a
  cross-sectional study}.  {\it BMJ Open. }2014;4:e004958.
\newblock \url{http://dx.doi.org/10.1136/bmjopen-2014-004958}. Accessed
  November 30, 2016.

\bibitem{Russ1980}
Russ SB. {A translation of Bolzano's paper on the Intermediate Value Theorem}.
  {\it Historia Mathematica. }1980;7(2):156--185.

\bibitem{Burden2011}
Burden RL, Faires JD. {\it {Numerical Analysis}}.
\newblock Boston: Brooks/Cole, Cengage Learning; 9~ed.2010.

\bibitem{Nemes2009}
Nemes S, Jonasson JM, Genell A, Steineck G. {Bias in odds ratios by logistic
  regression modelling and sample size}.  {\it BMC Medical Research
  Methodology. }2009;9(1):56.

\bibitem{Burton2006}
Burton A, Altman DG, Royston P, Holder RL. {The design of simulation studies in
  medical statistics}.  {\it Statistics in Medicine. }2006;25:4279--4292.

\bibitem{White2010a}
White IR. {simsum: Analyses of simulation studies including Monte Carlo error}.
   {\it The Stata Journal. }2010;10(3):369--385.

\bibitem{StataCorp2015b}
StataCorp. {\it {Stata statistical software: release 14}. } 2015.

\bibitem{Royston2004}
Royston P. {Multiple imputation of missing values}.  {\it The Stata Journal.
  }2004;4(3):227--241.

\bibitem{White2010b}
White IR, Carlin JB. {Bias and efficiency of multiple imputation compared with
  complete-case analysis for missing covariate values}.  {\it Statistics in
  Medicine. }2010;29:2920--2931.

\bibitem{Bartlett2015}
Bartlett JW, Harel O, Carpenter JR. {Asymptotically unbiased estimation of
  exposure odds ratios in complete records logistic regression}.  {\it American
  Journal of Epidemiology. }2015;182(8):730--736.

\bibitem{IMSHealth2015}
{IMS Health}. {Our data}  \url{http://www.epic-uk.org/our-data/our-data.shtml}.
  Accessed November 30, 2016.

\bibitem{Blak2011}
Blak BT, Thompson M, Dattani H, Bourke A. {Generalisability of The Health
  Improvement Network (THIN) database: demographics, chronic disease prevalence
  and mortality rates}.  {\it Informatics in Primary Care. }2011;19:251--255.

\bibitem{Bourke2004}
Bourke A, Dattani H, Robinson M. {Feasibility study and methodology to create a
  quality-evaluated database of primary care data}.  {\it Informatics in
  Primary Care. }2004;12:171--177.

\bibitem{Chisholm1990}
Chisholm J. {The Read clinical classification}.  {\it BMJ.
  }1990;300(6732):1092.

\bibitem{Dave2009}
Dave S, Petersen I. {Creating medical and drug code lists to identify cases in
  primary care databases}.  {\it Pharmacoepidemiology and Drug Safety.
  }2009;18:704--707.

\bibitem{Townsend1988}
Townsend P. {\it {Health and deprivation: inequality and the north}}.
\newblock London: Croom Helm; 1988.

\bibitem{Maguire2009}
Maguire A, Blak BT, Thompson M. {The importance of defining periods of complete
  mortality reporting for research using automated data from primary care}.
  {\it Pharmacoepidemiology and Drug Safety. }2009;18:76--83.

\bibitem{Horsfall2013}
Horsfall L, Walters K, Petersen I. {Identifying periods of acceptable computer
  usage in primary care research databases}.  {\it Pharmacoepidemiology and
  Drug Safety. }2013;22:64--69.

\bibitem{Sharma2016a}
Sharma M, Petersen I, Nazareth I, Coton SJ. {An algorithm for identification
  and classification of individuals with type 1 and type 2 diabetes mellitus in
  a large primary care database}.  {\it Clinical Epidemiology.
  }2016;8:373--380.
\newblock \url{https://doi.org/10.2147/CLEP.S113415}. Accessed November 30,
  2016.

\bibitem{Sharma2016b}
Sharma M, Nazareth I, Petersen I. {Trends in incidence, prevalence and
  prescribing in type 2 diabetes mellitus between 2000 and 2013 in primary
  care: a retrospective cohort study}.  {\it BMJ Open. }2016;6(1):e010210.
\newblock \url{http://dx.doi.org/10.1136/bmjopen-2015-010210}. Accessed
  November 30, 2016.

\bibitem{OfficeforNationalStatistics2012}
{Office for National Statistics}. {Ethnicity and national identity in England
  and Wales: 2011}
  \url{http://www.ons.gov.uk/peoplepopulationandcommunity/culturalidentity/ethnicity/articles/ethnicityandnationalidentityinenglandandwales/2012-12-11}.
  Accessed February 19, 2018.

\bibitem{Cornish2015}
Cornish R, Tilling K, Boyd A, Macleod J, {Van Staa} T. {Using linkage to
  electronic primary care records to evaluate recruitment and nonresponse bias
  in the Avon Longitudinal Study of Parents and Children}.  {\it Epidemiology.
  }2015;26(4):e41--2.

\bibitem{Kenward2007}
Kenward MG, Carpenter J. {Multiple imputation: current perspectives}.  {\it
  Statistical Methods in Medical Research. }2007;16:199--218.

\bibitem{Wood2004}
Wood AM, White IR, Thompson SG. {Are missing outcome data adequately handled? A
  review of published randomized controlled trials in major medical journals}.
  {\it Clinical Trials. }2004;1:368--376.

\bibitem{HayatiRezvan2015}
Rezvan PH, Lee KJ, Simpson JA. {The rise of multiple imputation: a review of
  the reporting and implementation of the method in medical research}.  {\it
  BMC Medical Research Methodology. }2015;15:30.

\bibitem{UKDataService}
{UK Data Service}. {Health Survey for England}
  \url{https://discover.ukdataservice.ac.uk/series/?sn=2000021}. Accessed
  November 30, 2016.

\bibitem{Raghunathan2015}
Raghunathan T. {\it {Missing data analysis in practice}}.
\newblock Boca Raton: Chapman {\&} Hall/CRC; 2015.

\end{thebibliography}

%
\section*{Supporting information}
\appendix
\counterwithin{table}{section}
\counterwithin{figure}{section}

\section{Weighted multiple imputation for a binary/categorical covariate} 
 The procedure of the weighted multiple imputation is as follows. In the imputation step, weights derived from the population marginal distribution of the incomplete variable are attached to the complete records, and a weighted (multinomial) logistic regression model is fitted to the complete records to obtain the maximum likelihood estimates of the imputation model's parameters $\widehat{\boldsymbol{\theta}}$ and their asymptotic sampling variance $\widehat{\boldsymbol{U}}$. New parameters are then drawn from the large-sample normal approximation $N(\widehat{\boldsymbol{\theta}}, \widehat{\boldsymbol{U}})$ of its posterior distribution, assuming non-informative priors. Finally, imputed values are drawn from the (multinomial) logistic regression using these new parameters. Note that \textit{no weights} are used when fitting the substantive scientific model to the imputed data. 

\subsection{Derivation of the marginal weights} 
The idea of augmenting the standard MI method with weights is related to the technique of post-stratification weighting, which is commonly used in survey non-responses when the population distributions are known.\cite{Raghunathan2015} To post-stratify the sample, weights are calculated to bring the sample distribution in line with the population. Suppose that in a survey, one of the variables measured is ethnicity, which is categorised into four groups (White, Black, Asian, and Other). If the population distribution of ethnicity is available, the distribution of ethnicity among survey respondents can be compared with the population distribution. Suppose that a proportion $p^{\text{obs}} = 0.8$ of the survey respondents give their ethnicity as White, whereas the population has $p^{\text{pop}} = 0.6$ in this category. The White category is over-represented in the survey respondents, but can be made representative of the population by assigning to the responses a post-stratification weight $w^{\text{ps}} < 1$, such that
\begin{equation*} 
        w^{\text{ps}} = 1/(p^{\text{obs}}/p^{\text{pop}}) = 1/(0.8/0.6)=0.75.
\end{equation*}
In adapting this idea to MI, we need to address the complication arising because the \textit{completed} data obtained after MI consist of both observed and imputed (missing) data. Naive use of post-stratification weights in MI will recover the correct population distribution in the imputed data. However, since the observed data remain the same, the distribution in the completed data will not be matched to that in the population. Therefore, some \textit{compensation} for the lack of representativeness in the observed data is needed in the imputed data so that the correct population distribution can be recovered after imputation. Continuing with the survey example, suppose that we survey $200$ individuals, $100$ of whom respond with their ethnicity. A proportion $p^{\text{obs}} = 0.8$ of these $100$ responses are in the White group. If the population proportion of this group is $p^{\text{pop}} = 0.6$, we would expect to have $120$ White individuals in the survey sample. This implies that among the $100$ individuals with missing ethnicity, we need to impute ethnicity of $40$ individuals as White, i.e. the proportion of the White category required in the missing data, $p^{\text{req}}$, is equal to $0.4$. To make the completed (observed and imputed) data of this category representative of the population, we need to weight respondents of this category in the imputation model by
\begin{equation*}
    1/(p^{\text{obs}}/p^{\text{req}}) = 1/(0.8/0.4) = 0.5,
\end{equation*}
\noindent
which is smaller than the corresponding naive post-stratification weight above, since it compensates for the over-representation among the survey respondents of White ethnicity.
\par
More generally, suppose that we seek to collect a $J$-level variable $x$ in a sample of size $n$, resulting in $x$ being observed for $n^{\text{obs}}$ subjects and missing for $n^{\text{mis}}$ subjects, $n^{\text{obs}}+n^{\text{mis}}=n$. Let $p_{j}^{\text{obs}}$ and $p_{j}^{\text{req}}$ denote the level-$j$ proportions of $x$ in the observed and imputed data respectively, such that $p_{j}^{\text{obs}}n^{\text{obs}}= n_{j}^{\text{obs}}$, and $p_{j}^{\text{req}}n^{\text{mis}}= n_{j}^{\text{req}}$, where $j = 1, \ldots, J$. Let $p_{j}^{\text{pop}}$ denote the level-$j$ proportion of $x$ in the population, which is assumed to be known. The aim here is to find $p_{j}^{\text{req}}$ for each level of $x$ such that the number of subjects in the completed data after imputation is equal to the expected number implied by the corresponding population proportion, i.e. $n_{j}^{\text{obs}}+n_{j}^{\text{req}}=p_{j}^{\text{pop}}n$. 
The level-$j$ proportion of $x$ required in the imputed data, $p_{j}^{\text{req}}$, is given by
\begin{equation*}
        p_{j}^{\text{req}}=\frac{p_{j}^{\text{pop}}n - p_{j}^{\text{obs}} n^{\text{obs}}}{n^{\text{mis}}}.
\end{equation*}
\noindent
Therefore, the weight for group $j$, which we refer to as the `marginal weight' and denote by $w_{j}^{\text{m}}$, is
\begin{equation*}
        w_{j}^{\text{m}}=1/(p_{j}^{\text{obs}}/p_{j}^{\text{req}}).
\end{equation*}

\subsection{Derivation of the conditional weights}
The marginal weights introduced above only depend on the population distribution of the incomplete variable. However, if there are (fully observed) covariates in the imputation model, the associations between these variables and the incomplete variable distribution are not reflected in such weights. We therefore adjust the marginal weights to obtain another set of weights, termed the `conditional weights', which account for covariates in the imputation model. These weights are derived using the marginal distribution of the incomplete variable obtained after having estimated the parameters of an imputation model assuming MAR in the complete records. Suppose that an imputation model is fitted to the complete records, and the corresponding predicted probabilities of the incomplete variable (averaged over the covariates) are obtained and applied to the missing data. Let ${p}_{j}^{\text{pred}}$ denote the resulting predicted level-$j$ proportion of $x$ in the completed data, then the level-$j$ proportion required in the imputed data is given by
\begin{equation*}
        p_{j}^{\text{req}} = \frac{p_{j}^{\text{pop}} n - {p}_{j}^{\text{pred}} n^{\text{obs}}}{n^{\text{mis}}},
\end{equation*}
\noindent
and the conditional weight for group $j$, denoted by $w_{j}^{\text{c}}$ , is
\begin{equation*}
        w_{j}^{\text{c}}=1/({p}_{j}^{\text{pred}}/p_{j}^{\text{req}}).
\end{equation*}
\par
In this approach, the effects of covariates in the imputation model are reflected in the predicted probabilities ${p}_{j}^{\text{pred}}$, which are then used to derive the conditional weights for weighted MI. 

\section{Analytic study -- bias in a $2 \times 2$ contingency table}
\label{app2}

In the $2\times 2$ contingency table of a complete binary outcome variable $y$ and an incomplete binary covariate $x$ (\cref{sec2}), we calculate analytic bias in the analysis model's parameter estimates (defined as $\hat{\beta} - \beta$) after missing values in $x$ are handled by (i) a CRA, (ii) standard MI, (iii) marginal weighted MI, and (iv) conditional weighted MI. The analytic calculations are then verified by simulating a full-data sample with $n=10\,000$ observations of $x$ and $y$ from the following model
\begin{align*}
    & x \sim \text{Bernoulli}\left(p_{x}^{\text{pop}} = 0.7\right);\\
    &\text{logit}\left[p\left(y=1 \mid x\right)\right] = \beta_{0} +\beta_{x}x,
\end{align*}
where $\beta_{0}=\text{ln}\left(0.5\right)$ and $\beta_{x}=\text{ln}\left(1.5\right)$. Missing values in $x$ are generated using selection models M1--M4 with a range of values for the selection parameters $\alpha$ (Table \ref{tab:anstudy_selectionparam}).
\begin{table}[t!]
	\renewcommand{\arraystretch}{1}
	\setlength{\tabcolsep}{5pt}
	\centering
	\caption{Analytic study: values of selection parameters for generating missingness in $x$ used in simulations conducted to verify analytic calculations.}
	\label{tab:anstudy_selectionparam}
	\begin{tabular}{cccccc}
	\toprule
\multirow{2}{*}{\begin{tabular}[c]{@{}c@{}}Missingness \\ model\end{tabular}} & \multirow{2}{*}{\begin{tabular}[c]{@{}c@{}}Linear predictor of selection model\\ $\text{logit}\left[p\left[(r=1 \mid x, y\right)\right]$\end{tabular}} & \multicolumn{3}{c}{Selection parameter}                         & \multirow{2}{*}{\begin{tabular}[c]{@{}l@{}}\% missing $x$\end{tabular}} \\\cmidrule(l{2pt}r{2pt}){3-3}\cmidrule(l{2pt}r{2pt}){4-4}\cmidrule(l{2pt}r{2pt}){5-5}
                                                                              &                                                                                                                                                        & $\alpha_{0}$        & $\alpha_{x}$        & $\alpha_{y}$        &                                                                                   \\ \midrule
M1                                                                            & $\alpha_{0}$                                                                                                                                           & $\left[-3,3\right]$ &                     &                     & $5$--$95$                                                                          \\
M2                                                                            & $\alpha_{0} + \alpha_{y}y$                                                                                                                             & $\left[-3,3\right]$ &                     & $\left[-3,3\right]$ & $3$--$97$                                                                            \\
M3                                                                            & $\alpha_{0} + \alpha_{x}x$                                                                                                                             & $\left[-3,3\right]$ & $\left[-3,3\right]$ &                     & $2$--$98$                                                                            \\
M4                                                                            & $\alpha_{0} + \alpha_{x}x + \alpha_{y}y$                                                                                                               & $0.5$                 & $\left[-3,3\right]$ & $\left[-3,3\right]$ & $9$--$84$                                                                            \\ \bottomrule
	\end{tabular}
\fnote{Note: $r$: response indicator of $x$.}
\end{table}
\par
\begin{figure}[htbp!]
    \centering
   	\caption{Analytic study: analytic bias when $x$ is MAR conditional on $y$ (M2).}
   	\label{fig:biascal_mary}
    	\includegraphics[scale=0.55]{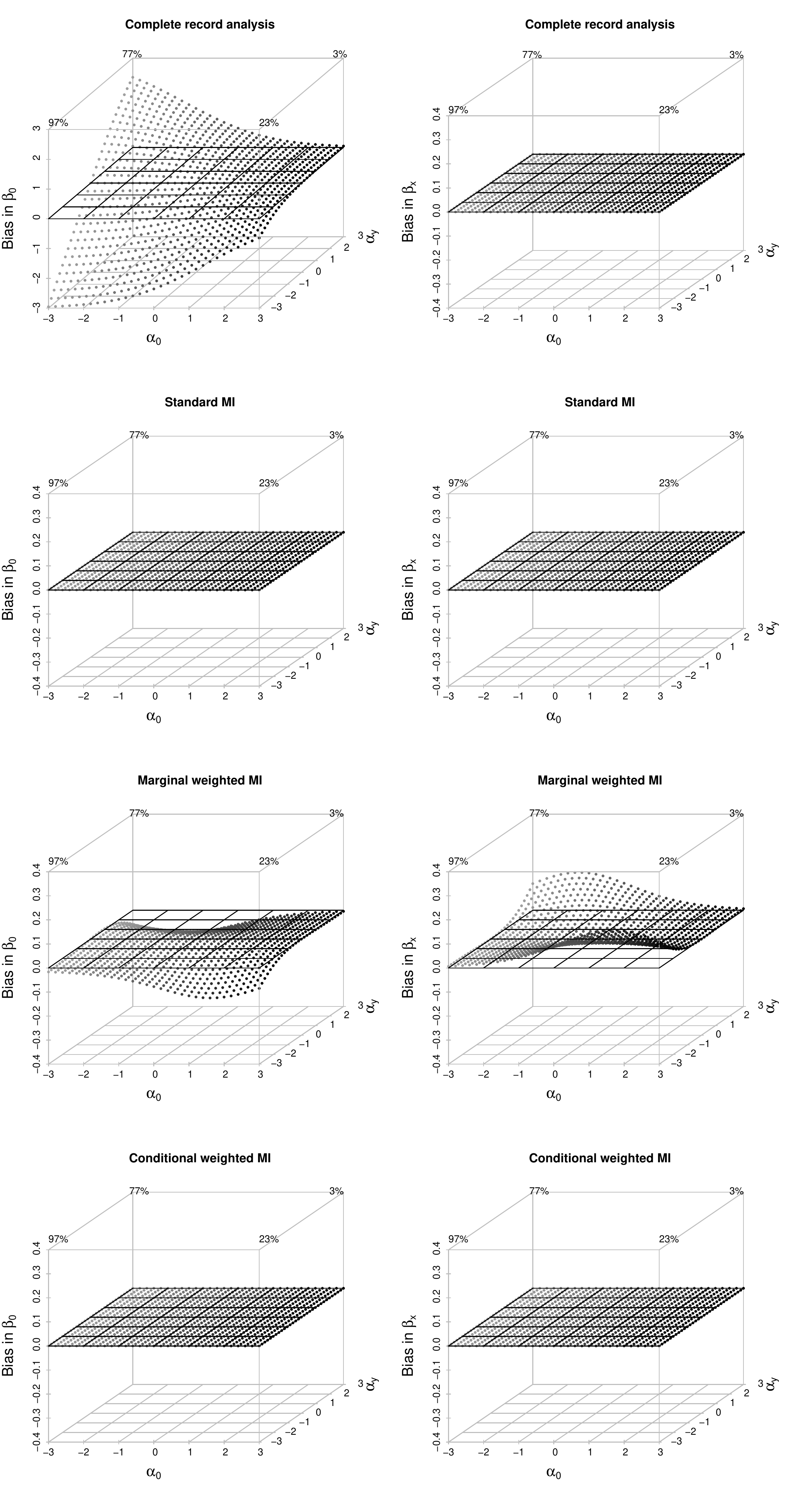}
    	\fnote{Note: selection parameters $\alpha_{0} \in [-3,3]$, $\alpha_{y} \in [-3,3]$; corresponding percentages of missing $x$ are presented for extreme values ($\pm 3$) of the $\alpha$ parameters.}
\end{figure}
\begin{figure}[htbp!]
    \centering
   	\caption{Analytic study: analytic bias when $x$ is MNAR dependent on $x$ (M3).}
	\label{fig:biascal_mnarx}
    	\includegraphics[scale=0.55]{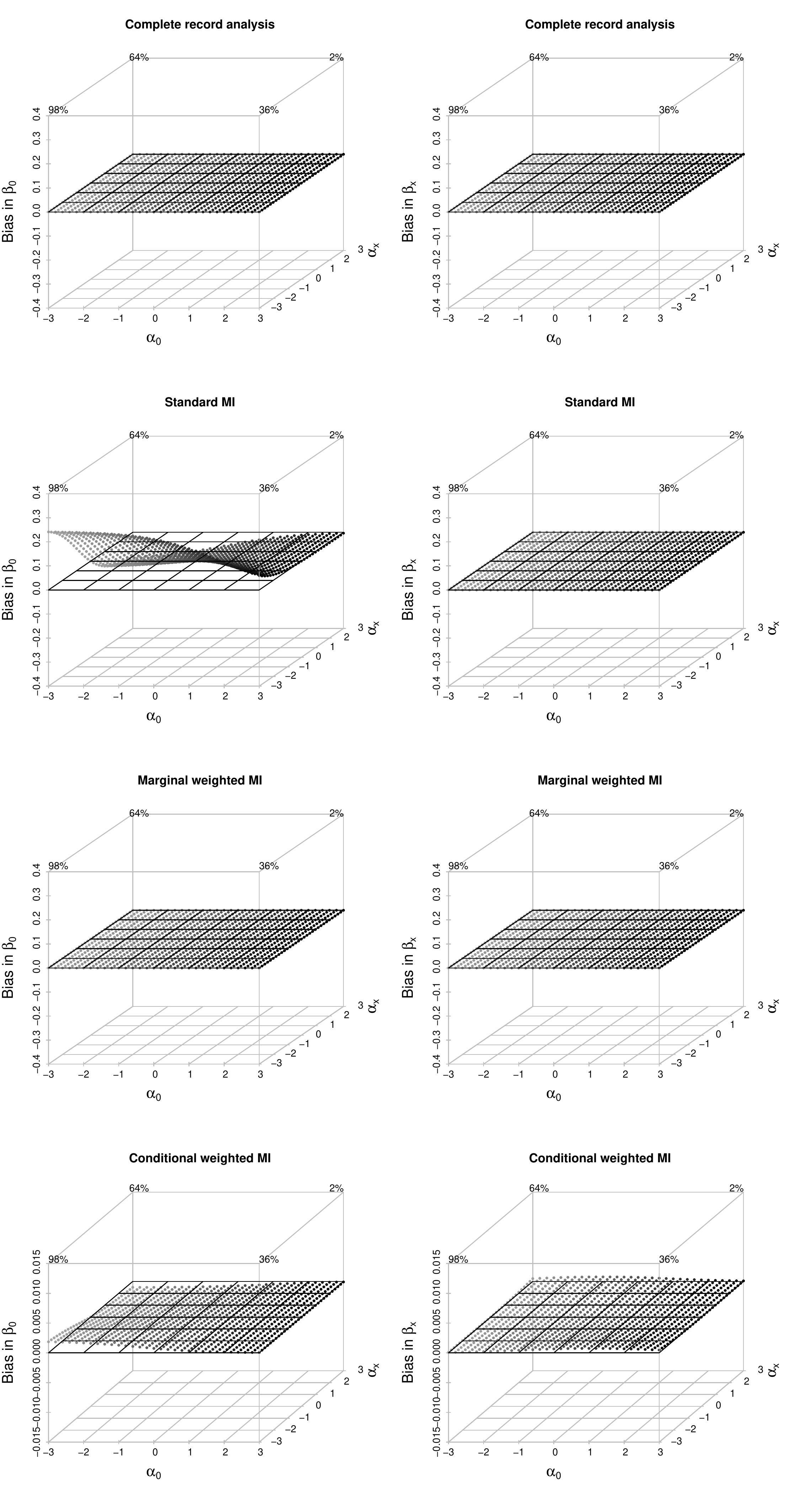}
    	\fnote{Note: selection parameters $\alpha_{0} \in [-3,3]$, $\alpha_{x} \in [-3,3]$; corresponding percentages of missing $x$ are presented for extreme values ($\pm 3$) of the $\alpha$ parameters.}
\end{figure}
\begin{figure}[htbp!]
   	\centering
   	\caption{Analytic study: analytic bias when $x$ is MNAR dependent on $x$ and $y$ (M4).}
	\label{fig:biascal_mnarxy}
    	\includegraphics[scale=0.55]{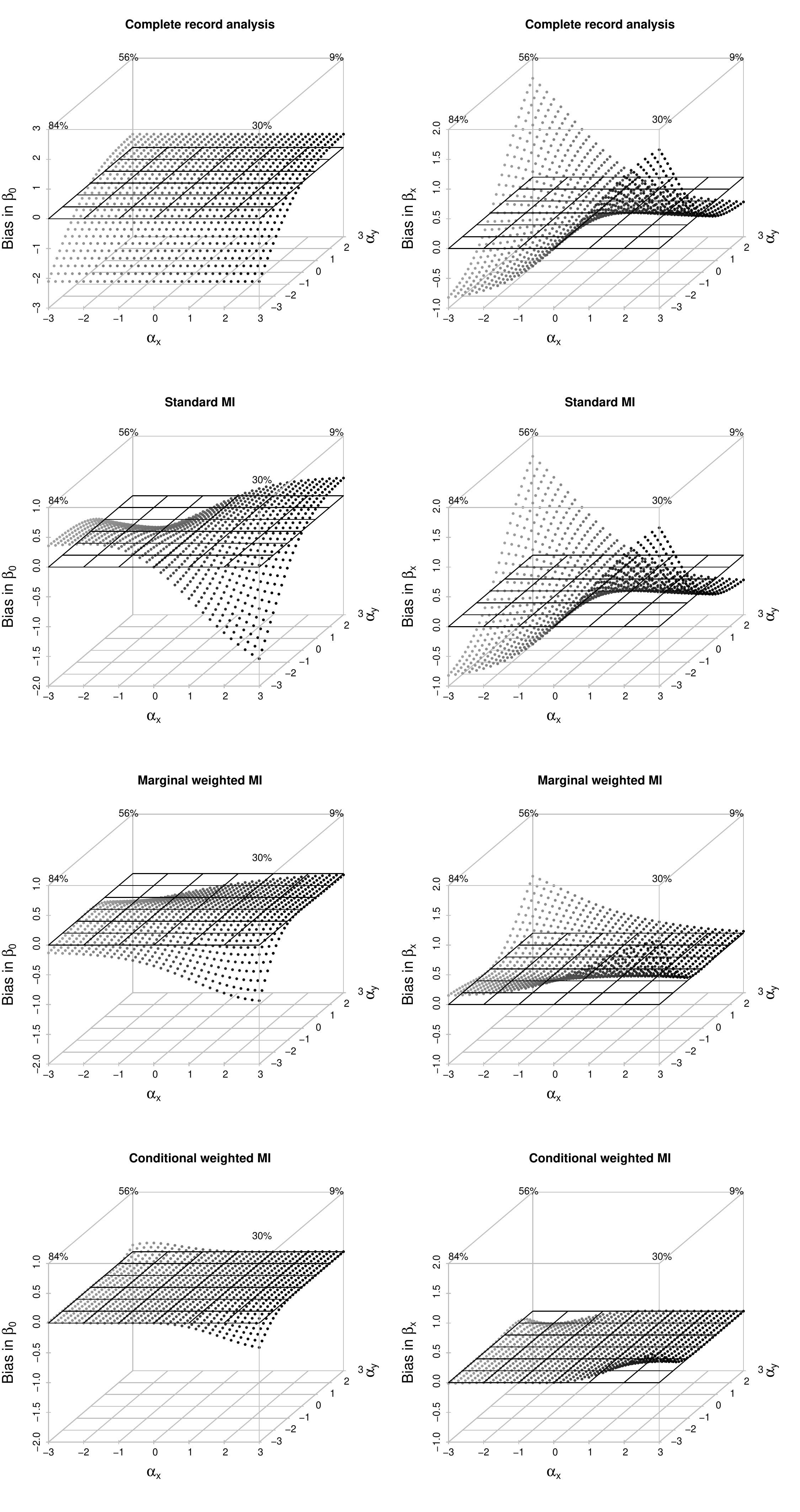}
    	\fnote{Note: selection parameters $\alpha_{0} = 0.5, \alpha_{x} \in [-3,3]$, $\alpha_{y} \in [-3,3]$; corresponding percentages of missing $x$ are presented for extreme values ($\pm 3$) of the $\alpha$ parameters.}
\end{figure}

Figure \cref{fig:biascal_mary,fig:biascal_mnarx,fig:biascal_mnarxy} present the analytic bias in CRA, standard MI, marginal and conditional weighted MI under MAR and MNAR mechanisms with the various values of the selection parameters. When $x$ is MCAR (M1), all methods provide unbiased parameter estimates, as suggested by the calculations (results not shown). 
\par
When $x$ is MAR conditional on $y$ (M2, Figure \ref{fig:biascal_mary}), standard MI and conditional weighted MI are unbiased, while bias is observed for CRA in $\beta_{0}$, and for marginal weighted MI in both parameter estimates. This bias is due to the marginal weights not accounting for the association between $x$ and $y$ in the imputation model for $x$. As a result, marginal weights do not successfully recover the correct distribution of $x$ after MI. 
\par
Both parameter estimates are unbiased in marginal weighted MI when $x$ is MNAR dependent on $x$ (M3, Figure \ref{fig:biascal_mnarx}), while standard MI leads to noticeable bias in the estimate of $\beta_{0}$. Bias in conditional weighted MI is small and occurs for extreme values of the selection parameters. Since missingness in $x$ does not depend on $y$ under M3, CRA is unbiased in both parameter estimates as the theory predicts. 
\par
Under the last missingness mechanism when $x$ is MNAR dependent on both $x$ and $y$ (M4, Figure \ref{fig:biascal_mnarxy}), none of the methods result in unbiased parameter estimates. However, bias appears to be the smallest in conditional weighted MI. Although bias is present in both standard MI and marginal weighted MI, the magnitude of bias is smaller in marginal weighted MI compared to standard MI. Under this missingness mechanism, conditional weighted MI can be regarded as a hybrid of marginal weighted MI and standard MI. The conditional weights correct for some bias introduced by $x$ in the selection model in a similar manner to the marginal weights under M3; the method also alleviates some residual bias similarly to standard MI under M2. 
\par
Overall, these results suggest that under the missingness mechanisms considered in this paper, calibrated-$\delta$ adjustment MI provides a more general solution for accommodating missing data in $x$ and is therefore the preferred method compared to standard MI and marginal and conditional weighted MI.

\end{document}